\documentstyle[11pt,aaspp4,flushrt]{article}

\def\fsl{\mbox{R$_{\rm f}$}}

\def\kev{\mbox{keV}}

\def\kms{\mbox{km s$^{-1}$}}
\def\kpch{\mbox{$h^{-1}$kpc}}
\def\mpc{\mbox{Mpc}}
\def\mpch{\mbox{$h^{-1}$Mpc}}
\def\msun{\mbox{M$_\odot$}}     
\def\msunh{\mbox{$h^{-1}$M$_\odot$}}

\def\omew{\mbox{$\Omega_{WDM}$}}

\def\mw{\mbox{m$_{W}$}}
\def\modI{\mbox{$\Lambda$WDMt2}}
\def\modII{\mbox{$\Lambda$WDM}}
\def\modIII{\mbox{$\Lambda$WDM$_{7.5}$}}
\def\modIV{\mbox{$\Lambda$WDM$_{15}$}}
\def\modV{\mbox{$\Lambda$WDM$_{30}$}}
\def\modVI{\mbox{$\Lambda$WDM$_{60}$}}
\def\modVII{\mbox{$\Lambda$CDM$_{60}$}}
\def\modVIII{\mbox{$\Lambda$WDM$_{60}$t2}}
\def\modIX{\mbox{$\Lambda$WDM$_{60}$t4}}
\def\modX{\mbox{$\Lambda$WDM$_{60}$t16}}
\def\mfs{\mbox{M$_{\rm f}$}}
\def\mvir{\mbox{M$_{\rm v}$}}
\def\rv{\mbox{r$_{\rm v}$}}
\def\rc{\mbox{r$_{\rm c}$}}
\def\vth{\mbox{v$_{\rm th}$}}
\def\vthw{\mbox{v$_{\rm th}^{\rm w}$}}
\def\vrms{\mbox{v$_{\rm rms}$}}

\def\mathnew{\mathsurround=0pt}
\def\ref{\par\noindent\hangindent=2pc \hangafter=1 }
\def\simov#1#2{\lower .5pt\vbox{\baselineskip0pt
    \lineskip-.5pt\ialign{$\mathnew#1\hfil##\hfil$\crcr#2\crcr\sim\crcr}}}  
\def\simgreat{\mathrel{\mathpalette\simov >}}

\def\'#1{\ifx#1i{\accent"13\i}\else{\accent"13#1}\fi}
\def\eg{e.g.,}

\def\Colinetal{Paper I}

\begin{document}
 
\title{Formation and structure of halos in a warm dark matter
cosmology}

\author{Vladimir Avila-Reese and Pedro Col\'in}
\affil{Instituto de Astronom\'{\i}a, U.N.A.M., A.P. 70-264, 04510, M\'exico, 
D.F., M\'exico}

\author{Octavio Valenzuela}
\affil{Astronomy Department, New Mexico State University, Box 30001, Department
4500, Las Cruces, NM 88003-0001, USA}

\author{Elena D'Onghia}
\affil{Universit\'{a} degli Studi di Milano, via Celoria 16,
 20100 Milano, Italy}

\author{and}

\author{Claudio Firmani}
\affil{Instituto de Astronom\'{\i}a, U.N.A.M., A.P. 70-264, 04510, M\'exico, 
D.F., M\'exico}

\begin{abstract}

Using high-resolution cosmological N-body simulations, we study
how the density profiles of dark matter halos are affected
by the filtering of the density power spectrum below a
given scale length and by the introduction of a thermal
velocity dispersion. In the warm dark matter (WDM) scenario,
both the free-streaming scale, \fsl, and the velocity dispersion, 
\vthw, are determined by the mass \mw\ of the WDM particle. We found 
that \vthw\ is too small to affect the density profiles
of WDM halos. Down to the resolution attained in our simulations
($\sim0.01$ virial radii), there is not any significant difference 
in the density profiles and concentrations of halos obtained in 
simulations with and without the inclusion of \vthw.
Resolved soft cores appear only when we increase 
artificially the thermal velocity dispersion to a value which is 
much higher than \vthw. We show that the size of soft cores in a 
monolithic collapse is related to the tangential velocity 
dispersion. The density profiles of the studied halos with 
masses down to $\sim0.01$ the filtering mass \mfs\ can be
described by the Navarro-Frenk-White shape; soft cores 
are not formed. Nevertheless, the concentrations of these halos 
are lower than those of the CDM counterparts and are approximately 
independent of mass. The cosmogony of halos with masses $\lesssim\mfs$ 
is not hierarchical: they form through monolithic collapse and by 
fragmentation of larger structures. The formation epoch of these 
halos is slightly later than that of halos with masses $\approx\mfs$. 
The lower concentrations of WDM halos with respect to their CDM 
counterparts can be accounted for their late formation epoch.   

Overall, our results point to a series of advantages of a WDM
model over the CDM one. In addition to solving the substructure problem,
a WDM model with $\fsl \sim 0.16$ Mpc ($\mw \approx 0.75\ \kev$;
flat cosmology with $\Omega_{\Lambda}=h=0.7$) also predicts
concentrations, a Tully-Fisher relation, and formation epochs
for small halos which seems to be in better agreement 
with observations relative to CDM predictions.     

\end{abstract}



\section{Introduction}


The damping of small-scale modes in the power spectrum
of density fluctuations has been proposed to overcome 
potential observational difficulties at small scales of the 
hierarchical cold dark matter (CDM) scenario for structure 
formation (\eg\ \cite{AFH98}; \cite{Moore99b}; \cite{SomDol00}; 
\cite{Hogan99}; \cite{KL99}; \cite{HD2000}; \cite{WC2000};
\cite{CAV}). The damping in the power spectrum may be produced 
either during its primordial generation (in the biased scalar 
inflationary models, for example) or due to Landau and 
free-streaming damping, which erase galactic or sub-galactic 
density fluctuations when the dark matter particles are 
warm (e.g., \cite{BPP82}); although the free-streaming 
damping could also be present in CDM models if CDM particles 
are non-thermally produced (\cite{Lin00}). In this paper we 
study warm dark matter (WDM) models; however, it is 
important to remark that our results are also valid for other
models where the power spectrum is damped at small scales. 

A potential shortcoming of the CDM scenario,  is the large amount
of substructure (satellites) that is predicted for
Milky-Way like halos with respect to observations under the
assumption that at each small halo a dwarf galaxy is formed
(\cite{Kauffmann93}; Klypin et al. 1999; \cite{Moore99a}; but see 
e.g., \cite{BKW2000} for an alternative solution). 
The cosmological N-body experiments 
of Col\'{\i}n et al. (2000, hereafter \Colinetal) have 
shown that the observed satellite circular velocity function 
of Milky Way and Andromeda can be reproduced
if the power spectrum is exponentially damped (filtered) at scales 
$\sim 0.1-0.2$ Mpc (a flat cosmological model with 
$\Omega_\Lambda = 0.7$ and $h = 0.7$ was used).  
A free-streaming scale \fsl\ of $0.1 - 0.2\ \mpc$ is attained for particle 
masses \mw\ of $\sim 0.6 - 1\ \kev$ (\cite{SomDol00}).
Interestingly, \cite{NSDCh2000} derived a lower limit for \mw\ of 750 eV from 
the restriction that the predicted power spectrum should
reproduce the observed properties of the Lyman$-\alpha$ forest 
in quasar spectra. 

In \Colinetal\ it was also found that the concentrations of satellite 
WDM halos
decrease as \fsl\ increases, being these concentrations 
lower than those of the corresponding CDM halos. Unfortunately, 
these small halos were poorly resolved and we could not explore
in detail the inner density profiles of halos with masses below \mfs.
Studying the density profiles of WDM halos with high resolution
is necessary, in particular because another potential problem of 
the CDM scenario is related namely to the inner density profile and
concentration of small halos. It seems to be a discrepancy between 
numerical simulation results and observations of dwarf and 
low surface brightness (LSB) galaxies (\eg\ \cite{Moore94}; 
\cite{FP94}; \cite{Burkert95}; \cite{dBM97};
de Blok et al. 2001) although other
observational studies have challenged this result (van den Bosch
et al. 2000; Swaters, Madore \& Trewhella 2000; van den Bosch
\& Swaters 2000).  

In the present paper the density profiles of WDM halos with 
masses close to or up to $\sim 100$ times smaller than \mfs\ are 
studied with a mass resolution $2-3$ orders of magnitude larger
than in \Colinetal. Our conclusions will be based partly 
on simulations with a power spectrum corresponding to $\mw\sim 0.6$ KeV.
However, in order to study with high resolution the halos with masses below 
\mfs, we need to increase \mfs. Thus, we will also use simulations for 
$\mw\sim 0.13$ KeV ($\mfs\sim 10^{14} \msunh$) and assume that the 
structure of halos should depend on the halo mass-to-\mfs\ ratio rather 
than on the specific value of \mfs. The cosmogony of halos with 
masses $<\mfs$ is not hierarchical. Halos with masses close to or larger 
than \mfs\ by factors of $5-10$ assemble a large fraction of their 
present-day mass by a coherent and nearly monolithic collapse, whereas halos 
with masses smaller than \mfs\ form from the
gravitational fragmentation of larger pancake-like structures,
as in the top-down scenario (Zel'dovich 1970).
Firmani et al. (2000b) emphasized that the density profiles of
dark halos formed through a monolithic collapse depend significantly 
on the velocity dispersion of the collapsing particles (see also
Aguilar \& Merrit 1990); this
velocity dispersion may be primordial (thermal relict) or 
can be acquired during an inhomogeneous gravitational collapse
(dynamical). 

The non-zero thermal velocities of warm particles (warmons) 
should produce minimal shallow halo cores because of the phase 
packing limit (\eg\ Gunn \& Tremaine 1979). However, this 
thermal velocity dispersion, \vthw, for warmons with masses of 
$\sim 1\ \kev$ seems to be too small to produce a noticeable core
(Hogan \& Dalcanton 2000). In \Colinetal\ this velocity was 
neglected, leaving uncertain the effect it could have on the 
structure of simulated halos (Sellwood 2000). Here we will include 
 different thermal velocity values in the simulations in order
to explore their influence on the halo density profile. 
An analytical (dynamical) approach will be also used
to study the ``hot'' monolithic collapse and clarify the conditions
required to form shallow cores.

After the completion of this paper, a similar paper by Bode, 
Ostriker, \& Turok (2000) appeared in the preprint list of Los Alamos
National Laboratory. Bode et al. study the cosmogony and structure
of halos for models with damped power spectra in a cosmological 
box. They also find that halos with masses below the filtering mass 
are less concentrated and that these halos form later, relative to CDM. 
In addition to this, they also describe the spatial distribution and 
the mass function of these small halos; in this sense, their study 
complement ours.  

In section \S 2 we present the WDM models to be explored in this paper. In 
\S 3 we  briefly describe the numerical simulations. In \S 4 we 
present the results from our experiments aimed to study the density 
profiles and concentrations of halos with masses below the mass 
corresponding to the filtering scale. The influence of the
velocity dispersion over the inner density profile of halos 
formed by monolithic collapse is explored in \S 5.1. In \S 5.2
we present results from our WDM cosmological simulations including
the thermal velocity dispersion of warmons.
In \S 6.1 we describe and discuss how structure formation 
proceeds in simulations with a power spectrum suppressed at small
scales. The viability of the WDM scenario is disscused in 
\S 6.2. Finally, our conclusions are given in \S 7. 


\section{Cosmological Models}


As in \Colinetal, here we also use the currently favored flat 
low-density  universe with $\Omega_\Lambda = 0.7$, 
$h = 0.7$, and $\sigma _8=1$, and instead of CDM we introduce
WDM. Two characteristics distinguish WDM from CDM:
the damping of the small-scale density fluctuations and the fact
that thermal velocity dispersion might not be negligible at the
time of structure formation. The
free-streaming length \fsl\ is related to the mass of the warmon through the
following equation (\eg\ \cite{SomDol00}):
\begin{equation}
\fsl = 0.2\ (\omew h^2)^{1/3}\ \left( \frac{m_W}{1 \kev} \right)^{-4/3} \mpc,
\end{equation}
\noindent where \omew\ is the contribution to the energy density of 
the universe from WDM. Modifications to eq. (1) exist; for example,
a larger coefficient of proportionality is given by Pierpaoli et al. (1997).
Using their coefficient, for a given \fsl, \mw\ should be roughly
two times larger than the values used here. 

We follow Sommer-Larsen \& Dolgov and define
a characteristic free-streaming wavenumber $k_f$ as the $k$ for which
the WDM transfer function (eq. [4]) is one half; that is, $k_f = 0.46 / \fsl$.
Once \fsl\ has been fixed (this is the ``true'' input parameter
in our simulations), a characteristic filtering mass \mfs\ can be
defined conditionally. Unfortunately, this has not been done in an unique way in 
the literature, introducing some confusion; its value can vary
by about three orders of magnitude from one definition to another. 
For instance, defining \mfs\ simply as $4 \pi / 3 \bar\rho_m \fsl^3$ gives
a value 318 times smaller than the value obtained with the definition
given by Sommer-Larsen \& Dolgov (2000), which we use in this paper: 
\begin{equation}
\mfs \equiv \frac{4 \pi}{3} \bar\rho_{m} \left( \frac{\lambda_f}{2} \right)^3,
\end{equation}
\noindent where $\lambda_f \equiv 2 \pi / k_f = 13.6 \fsl$. This definition
of $\mfs$ is justified by the numerical experiments of \cite{TC2000}.
They found that the initial density distribution
produced by a sharp cut in the power spectrum at $k_c =
2 \pi / r_s$ (the exponential drop in the power spectrum in our case
is at $k_c = 2 \pi / \lambda_f$) is similar to that one produced when the initial
density field is convolved with a top-hat window of smoothing
radius $\sim r_s/2$. The analysis of the halo properties below shows that
the mass at which the WDM halos begin to diverge from the CDM ones is just
around the mass given by eq. (2). 

We use the WDM power spectrum given in Bardeen et al. (1986):
\begin{equation}
P (k) = T^2_{WDM} (k) P_{CDM} (k),
\end{equation}
\noindent where the WDM transfer function is approximated by
\begin{equation}
T_{WDM} (k) = \exp\left[ -\frac{k\fsl}{2} - \frac{(k\fsl)^2}{2} \right]
\end{equation}
\noindent and $P_{CDM}$ is the CDM power spectrum given by
 \cite{KH97} (see \Colinetal).


\section{The Numerical Simulations}


The simulations were performed using the multiple-mass
scheme variant of the Adaptive Refinement Tree 
(ART) N-body code (\cite{KKK97}). The ART code achieves high 
spatial resolution by refining the base uniform grid in all
high-density regions with an automated refinement algorithm.
The multiple-mass scheme, described in detail elsewhere (Klypin et 
al. 2000; see also \Colinetal), is used to increase the mass and spatial 
resolution in few {\it selected} halos, which hereafter we will 
call hosts as in \Colinetal. However, while in \Colinetal\
the host halos were structures much larger than \mfs\ and selected
to be relatively isolated, here, in simulations with large \fsl, 
they are smaller than \mfs\ and can be embedded within 
filaments and can belong to groups at the intersection of 
filaments. On the other hand, in both CDM and WDM, we call guest 
halos the bound structures surviving within the virial radii of the
host halos. These definitions 
are more technical than conceptual; note that in the simulations
just mentioned above, the host halos contain guest halos, but on their
own they may be ``guests'' of other larger systems.     

We start our simulations by making a low mass resolution run
with $32^3$ or $64^3$ particles in a grid with $256^3$ cells in
which all particles have the same mass. We use these runs
to select (host) halos to be re-simulated with higher mass
and force resolution. 
We then identify all particles within $\sim 2$ virial radii and 
trace them back to get their Lagrangian positions at $z = 40$. 
For the cosmology used in this paper, the virial radius 
\rv\ is defined as the radius within which the average halo density 
is 334 times the background density, according to the spherical 
collapse model. Next, the initial distribution with particles 
with different masses is generated and the models are evolved to 
the present time with the multiple-mass variant of ART. Models 
with $\fsl = 0.2\ \mpc$ have four mass levels (particles with 
masses 1, 8, 64, 256 $\times m_p$) whereas models with 
$\fsl = 1.7\ \mpc$ have three levels. The mass resolution on the
finest mass level corresponds to a box of $256^3$ particles 
in both series of simulations.      

As in \Colinetal, the Bound Density Maxima (BDM) 
group finding algorithm was used to
locate the host halos in the low mass resolution run.
The BDM algorithm finds the positions of 
local maxima in the density field smoothed at the scale of interest 
and applies physically motivated criteria to test whether a group 
of particles is a gravitationally bound system. The BDM is 
also used to find the host and the guest halos in the multiple-mass
high resolution simulations. 

In Table 1 we present an overview of all the simulations used
in this paper. The formal force resolution shown in column 6 is 
the size of a cell in the finest refinement grid, and the 
mass per particle (column 4) is the mass for those particles 
which belong to the finest level of mass resolution. The halo 
density profile can be studied with high confidence only for 
radii larger than $\sim 4$ times the formal force resolution 
(Klypin et al. 2000) and containing within them more 
than 50-100 particles. 

The first set of runs in Table 1 was aimed at studying 
in detail the density profiles and concentrations of halos with 
virial masses close to or below the corresponding filtering
mass \mfs. Model \modII\ is the same simulation presented in 
\Colinetal\ for $\fsl=0.2$ \mpc\ but run in a smaller box,
$L_{\rm box}=7.5$ \mpch, in order to attain more mass resolution. 
The host halo studied in this run is $1.7$ times less massive 
than the same halo in the run with $L_{\rm box}=15 \mpch$.
\footnote{This is in fact what we find in all of our  WDM 
experiments: the mass of a given halo becomes smaller as 
$L_{\rm box}$ is reduced. However, this mass reduction is not 
dramatic, and we do not expect it will change our conclusions
about concentrations and density profiles of the halos, because 
they both are not typically very dependent on mass.} Unfortunately, 
in the simulation with the box size reduced, the most massive 
guest halos turned out to be small, containing less than 1000 particles. 

Since it is not easy to know the masses of the guest halos {\it a priori}, 
and since our aim is just to explore the density profiles of halos with
masses smaller than \mfs, a better strategy is to study halos in 
simulations where we select them {\it a priori} (hosts), with the desired
masses below \mfs. Although from the technical point of view of our approach
these halos are hosts, they may be well embedded within larger halos, 
i.e., they could be satellites (guests). We set $\fsl = 1.7\ \mpc$ which 
corresponds to $\mfs = 1.7 \times 10^{14}\ \msunh$, and re-simulate with
high resolution smaller and smaller host halos in runs with decreasing 
box sizes. For this model (\mw=125 eV), the structure formation process 
is close to the hot dark matter regime. Nevertheless, in the assumption 
that the formation and structure of halos depend on $\mvir/\mfs$ rather 
than the specific value of \mfs, this model will give guidance to the 
physics on smaller scales in a model with a more realistic warmon mass. 
We think this assumption is plausible. In any case, general trends 
can be compared with the (low resolution) results from simulations with 
\mw=605 eV.

Runs \modIII, \modIV, \modV, and  \modVI\ (Table 1), are for 
$L_{\rm box}= 7.5, 15, 30$ and 60 \mpch, respectively
For comparison, we also run a CDM simulation ($\fsl = 0$)
with the same initial condition as \modVI\ (run \modVII). 
In model \modVI\ we re-simulated four host halos with masses $2-4$ 
times smaller than \mfs. They were resolved with $\sim4 - 7 \times 10^4$ 
particles. In order to obtain halos with masses even $\sim10$ times smaller
but with the same mass resolution, we reduced the box size by a factor of two.
This is run \modV, where four halos with masses $22-33$ times smaller than
\mfs\ were re-simulated. In run \modIV, the only host halo re-simulated is
already 45 times smaller than \mfs\ and has $3.54 \times
10^5$ particles. The guest halos (those contained within the virial 
radii of hosts) in all of these simulations have 
masses smaller than $\sim0.01$ \mfs\ and are resolved with
$\lesssim 7000$ particles. Finally, in run \modIII\ the 
selected halo is 300 times less massive than \mfs\ and it is resolved
with $2.73 \times 10^5$ particles.  

At this point, it is important to note the effect that a reduction
of the box size has on the results from WDM simulations. In CDM 
simulations we do not expect that a reduction in the box size 
affects the internal structure of halos with radii much smaller 
than $L_{\rm box}$ (e.g.,  Frenk et al. 1988). However, in WDM 
simulations the loss of long wavelengths may critically affect the 
formation and the structure of halos if the fundamental wavelength
($= L_{\rm box}$) is close to or smaller than the characteristic 
filtering wavelength $\lambda_f$. When dealing with WDM power spectra, 
one should keep in mind that structures of scales smaller than 
$\sim\lambda_f$ arise from a transfer of power from large to small 
scales (see \S 6.1). Hence, if the large
scale modes are omitted, especially those around the peak, then 
the formation of small structures becomes affected. 
For CDM this is not a problem because small scales have ever 
more power than the large ones and the structure formation process
is dominated by the smallest scales.

Since the filtering wavelength for $\fsl = 1.7\ \mpc$ is $\lambda_f =
16 \mpch$ (see \S 2), run \modIV\ is at the limit of confidence while 
for run \modIII, the mode corresponding to the peak
of the WDM power spectrum is absent (the fundamental mode lies beyond the
peak). The density profile of the halo re-simulated in this peculiar
model (a narrow and exponentially damped power spectrum) indeed
strongly deviates from profiles of halos obtained in other WDM simulations
with larger box sizes as will be seen in \S 4.

The second set of simulations shown in Table 1 was aimed at studying the 
effect that a warmon non-zero thermal velocity, \vthw, has on 
the internal structure of simulated halos. From this study
we conclude that \vthw\ does not affect the inner density profile
of WDM halos, at least down to the scale at which we are confident 
of the resolution of our simulations ($\sim0.01$\rv). Run \modI\ is the same
as \modII\ ($\fsl=0.2$ Mpc) but introducing a thermal velocity
component twice larger than \vthw.
Runs \modVIII, \modIX, and \modX, are the same as \modVI\ ($\fsl=1.7$ \mpc) 
but introducing thermal velocities with amplitudes 2, 4 and 16
times larger than the corresponding \vthw.

In order to attain a visual impression of how structure formation
does proceed for models with a damped power spectrum, we present in Fig.
1a snapshots at diferent epochs of the run \modVI. The figure
snapshots show the distribution of particles inside a cube of 40 \mpch\ 
comoving on a side. The center of each cube is the
center of mass of the particles that were traced back to the selected 
epoch from a sphere of radius 1.5 \mpch\ centered in one of the host 
halos at $z = 0$ (Fig. 1c, see below). In Figure 1b we show the same as 
in Fig. 1a, but for the corresponding CDM model, run \modVII. The different 
formation histories of halos with masses $\approx \mfs$ or below in 
models with and without damped power spectra are highlighted 
in these two panels; in
particular, we notice that these halos (i) form later in model \modVI\
than in \modVII\ (the amount of substructure at $z \sim 2$ in model 
\modVI\ is reduced to a solitary sharp filament) and (ii) that they are
only located within filaments in the WDM case while in the CDM one,
they can also be found outside the filaments (see also Bode et al. 
2001). In Fig. 1c we show a zoom of the central region of plot 1a at
two times, $z=1$ and $z=0$. In the right panel all particles inside a 
sphere of 1.5 \mpch\ centered on a host halo at $z=0$ are plotted. The 
mass of this halo is about half the filtering mass 
($\mfs = 1.7\times \ 10^{14} \ \msunh$) and it was one of the first 
structures to collapse in the simulation. In the left panel, the particles
plotted in the right panel were traced back to $z=1$. Note that the
virialized halos seem to emerge from the filamentation and fragmentation
of panckakes. In any case, the detailed study of halo 
formation in WDM-like models is beyond the scope of this paper.

An effect that is becoming typical of cosmological high-resolution 
simulations  with  damped power spectra is the finding of relatively small 
halos regularly spaced in filaments. An example of this is shown in Fig.
1a at $a = 0.5$; see those knots that lie inside the filament which
points from right to left to the center of the cube. Several of these knots
were indeed selected by our halo finding algorithm as self-gravitating
structures. The space regularity, also seen in other of our simulations,
make us to suspect that the origin of these very small halos with respect to 
\mfs\ may be a numerical artifact attributed perhaps to finite grid effects.
A detailed study of this effect is highly desirable since it might
produce an steepening of the circular velocity function distribution of 
satellites at the small velocity end.


\section{Density profiles of halos with masses near or below
the filtering mass}


In \Colinetal\ host halos of a few $10^{12} \msunh$ were
re-simulated using the multiple-mass scheme in a box size
of 15 \mpch\ for a $\Lambda$CDM model with $\fsl=0.2$ \mpc. 
The mass per particle in the finest level of 
resolution corresponded to $1.66\times 10^7 \msunh$.
Thus, the guest halos reported in \Colinetal\ (with scales
below the filtering scale) were poorly resolved; they only had
a few $10^2$ particles, a number not high enough to study 
in detail their inner density profiles. 
As mentioned in $\S 3$, we also run the same simulation of 
\Colinetal\  but with the box size reduced by a factor
of two (run \modII). We find that the 
most massive guest halos are smaller than the most massive
ones from the 15 \mpch\ run, in such a way that they were 
resolved with not much more particles than in \Colinetal.  
Figure 2 (upper panel) shows the density profiles of the host and 
guest halos with more than 1000 particles (the latter were shifted 
vertically in order to avoid too much overlapping). The radius
of the innermost point is at least four times 
the formal force resolution and contains more than 50 particles  
within it (see \S 3; this criteria applies for all density profiles 
shown in this paper). 
For the host halo, the innermost radius is $4\times0.1$ \kpch, and for the
guest halos is $\approx 1.2$ \kpch. 
As one sees from Figure 2, the density profiles of guest halos
--- whose masses are $\sim0.01$ times smaller than the filtering 
mass $\mfs=2.8\times10^{11}$ \msunh ---  are well described 
by the Navarro, Frenk, \& White (1997, NFW) profile (dashed lines), 
from $\sim0.04$ to 1 \rv.

As described in \S 3, for a series of $\Lambda$WDM simulations 
with $\fsl=1.7 \mpc$ and different box sizes, we re-simulated
with high resolution selected halos (hosts) with masses smaller
than the corresponding $\mfs=1.7 \times 10^{14}\ \msunh$, 
obtaining in this way halos with a large number particles.
  
In Fig. 3 we show the density profiles of the host halos with
masses $2-4$ times smaller than \mfs\ from run \modVI\ (filled 
circles) and those obtained in the same simulation but with CDM 
instead of WDM (run \modVII, empty circles). Although both 
simulations started with identical phases (seeds) and large scale 
normalization, the CDM halos are slightly more massive than the 
corresponding WDM ones (by factors $\sim1.1-1.4$). The dashed 
lines are the NFW
fit to the plotted profiles. As can be seen from this figure, 
the NFW shape describes rather well the density profiles of both 
the WDM and CDM halos, for radius ranging from $\sim0.01$ to 1 \rv, 
although in the innermost parts, the slope tends to be steeper than
$r^{-1}$, in particular for the CDM halos (Moore et al. 1999a).
The accuracy of the fit can be estimated with the parameter
$D\equiv\chi^{2}/N$, where
$\chi^{2}=\sum_i^N [\log(\rho_i/\rho_{an})]^2$ ($\rho_i$ and $\rho_{an}$
are the measured and analytical values of the density, respectively)
is the quantity used for minimization of the fitting, and
$N$ is the number of radial bins (points to fit).
For the host halos in runs \modVI\ and \modVII, on average
$D\approx 6.5$\% and 5.5\%, respectively.

In order to explore the structure of halos with masses much smaller  
than \mfs, we have run simulations with the same
\fsl=1.7 \mpc\ but with box sizes of 30, 15 and 7.5 \mpch\ (runs \modV, \modIV,
and \modIII, respectively); in these simulations we re-simulate host 
halos with masses of $\approx 6-95\times 10^{11}\ \msunh$. Note
that these halos may be embedded within larger halos, i.e. they could
be satellites (see \S 3 for the technical definitions of host and
guest halos).

Figure 4 shows the density profiles of the four host halos
obtained in the run \modV\ (upper panel), and of the two host 
halos obtained in runs \modIV\ and \modIII\ (lower panel). 
For comparison, the density profile of a CDM halo of $2.5 \times
10^{12}\ \msunh$ presented in \Colinetal\ is also plotted (empty
circles). Though now the masses of the halos are up to $\sim$50
times smaller than \mfs, the NFW fit (dashed lines) continues to
be a good description for the density profiles of WDM halos
explored with accuracy down to $\sim0.01$ \rv. 
The deviation parameter $D$ is on average $\sim6$\% for host halos from 
run \modV\ and $\sim$3.5\% for the only host halo
from run \modIV. Nonetheless, the WDM halos are less 
concentrated and shallower in the center than their CDM counterparts. 

The density profile of the host halo from simulation \modIII\
shows a prominent shallow core and a disturbed outer density profile.
As this run has a box size already smaller than the filtering
wavelength $\lambda_f$, it samples only the high-frequency drop 
of the WDM power spectrum; we thus expect the inner structure of the 
formed halo to be significantly affected (see \S 3).

The results obtained here confirm our previous result: the WDM halos
with masses near or below the filtering mass are {\it less concentrated}
than the corresponding CDM halos (\Colinetal). In Figure 5a we plot
the c$_{1/5}$ concentration parameter\footnote{The concentration 
c$_{1/5}$ is defined as the ratio between the virial radius \rv\
and the radius where 1/5 of the total halo mass is contained 
(Avila-Reese et al. 1999). This definition of the concentration
parameter is independent of the particular fitting applied to the
halo density profile.} versus virial mass \mvir\ for the host 
halos from the different runs with $\fsl = 1.7\ \mpc$. We also 
include in this plot guest halos with more than 1000 particles. 
They are only a few and their inner density profiles are resolved
only down to $\sim0.03$ \rv.  The concentrations and inner density
profiles of surviving guest halos are not expected to be 
significantly affected by the fact that they are within larger
systems. In fact, some of the host halos in the different runs 
are also within larger halos. Results
for the WDM model with $\fsl = 0.2\ \mpc$ are also shown
in Fig. 5a: crosses are from \Colinetal\ and skeletal triangles
correspond to the \modII\ run presented here. Halos with less than 90 
particles were excluded. 

In order to compare the concentrations of WDM and CDM halos,
we also plot in Fig. 5a the results from our $\Lambda$CDM simulation of 
box size 60 \mpch\ (empty circles) as well as a linear 
fitting for thousands of isolated and clustered halos found in a
$\Lambda$CDM simulation (Avila-Reese et al. 1999). The trend in 
Fig. 5a is clear: as the halo mass becomes smaller
than \mfs, the c$_{1/5}$ concentration departs
more and more from the corresponding CDM concentration. Although
the dispersion in the concentration is big (see Fig. 8 in 
Avila-Reese et al. 1999 and Fig. 8 below)  
the trend seen in Fig. 5a for the WDM halos is clear and is 
out of the statistical dispersion of the c$_{1/5} - $\mvir\ relation 
of the CDM halos.
For completeness, the concentrations obtained when fitting the
density profiles of the halos to a NFW profile (c$_{\rm NFW}$) are also
presented in panel (b) of Fig. 5 (only halos with $D<10$\% were 
included). The four halos from the 60 \mpch\ CDM simulation, 
as well as a linear interpolation for the c$_{\rm NFW} - \mvir$ relation
obtained from the $\Lambda$CDM simulation presented in Avila-Reese et al. 
for halos whose density profiles are well described by 
the NFW shape, are also shown in this plot.

\section{Effect of thermal velocity dispersion on the 
density profile}

There are two relevant  conditions for the problem of halo
formation, namely the initial fluctuation density profile 
---which determines the kind of collapse the halo suffers--- 
and the amplitude of the (tangential) velocity dispersion \vrms\ 
of the collapsing particles. The CDM halos form hierarchically 
and the \vrms\ is acquired 
{\it dynamically} since early epochs due to interactions of 
substructures with the global tidal field.
In a WDM scenario, the first structures to collapse are those 
with masses close to \mfs; they assemble the mass \mfs\ almost 
synchronously followed by some mass accretion (quasi-monolithic 
collapse). It is possible that structures formed by fragmentation 
also suffer a quasi-monolithic collapse. As the collapse of halos 
less massive than \mfs\ is delayed and the amount of substructure 
is small (compared to halos formed hierarchically), one expects 
that particles will acquire a lower tangential velocity dispersion 
than in the hierarchical case, although this is a question to be 
explored in the numerical simulations. Instead, WDM  
particles have an intrinsec (residual) {\it thermal} velocity 
dispersion. 

An interesting question to explore is how the angular momentum of 
the particles --- due to either a residual thermal velocity or to
a \vrms\ acquired dynamically --- affects the inner structure of 
virializing systems. For halos formed hierarchically,
the velocity dispersion is not relevant for the formation
of a soft core (e.g, Avila-Reese et al. 1998; Huss, Jain, \& 
Steinmetz 1999), but for halos formed monolithically it may be. 
In \S 5.1 we present an {\it heuristic} (analytical) model 
of halo virialization which allows us to understand the 
role that velocity dispersion plays in a monolithic collapse. We 
will also find an expression to relate the soft core radius with
\vth, using N-body simulations for hot monolithic collapse
and virialization. In \S 5.2 
results from cosmological N-body simulations, including several
values of \vth, will be presented.

\subsection{Thermal monolithic collapse and soft cores}
Let us study the monolithic collapse of a sphere of mass $M$, radius 
$R_M$, and uniform density $\rho_M$ at maximum expansion. We can
introduce some thermal energy assuming that all particles 
move on elliptical orbits with a given eccentricity $e$. The 
constancy of the density and $e$ with radius asure that
all particles have the same orbital period $\tau$ and that 
the pericenter-to-apocenter ratio is the same for all of them,
respectively. Therefore, shell crossing is avoided.
For purely radial motion, 
$(e=1)$, the collapse reaches a singular point. For $e < 1$, 
at maximum concentration, i.e. when each particle is found at 
pericenter, the collapse leads to a uniform sphere with radius
\begin{equation} 
r_c=R_M (1-e)/(1+e). 
\end{equation}
The existence of this maximum concentration 
state is at the base of the existence of a soft core in the virialized 
halo. Tangential velocity at apocenter is:
\begin{equation} 
 v_t^2= \frac{r^2 4\pi G\rho_M (1-e)}{3},   
\end{equation}
while, at the maximum expansion, the total thermal energy $T$ and 
the potential energy $W$ are given by:\\
\begin{equation} 
T=  \frac{8 \pi^2 G \rho_M^2}{15}(1-e)R_M^5
\end{equation}
\begin{equation} 
W=\frac{16 \pi^2 G}{15} \rho_M^2 R_M^5,
\end{equation}
where the relationship between $e$, T and W is: $2T/W=1-e$.

 The simplest approximation to estimate the virialized
density at a given radius (density profile) is to assume a time
average density at the same radius. This approach is based
on the statistical hypothesis that the time average density is 
representative of the virialized halo density.        
Using the conventional motion parametric equations:
\begin{eqnarray} 
r=\frac{R_M}{(1+e)} (1-e \cos\theta) \nonumber \\
t=\sqrt{\frac{3}{4 \pi G \rho_M (1+e)^3}} (\theta- e \sin\theta),
\end{eqnarray}               
the time average density at a radius $r$ between
$r_c$  and $R_M$ is:
\begin{eqnarray} 
<\rho (r)>&=& \frac{\int_{0}^{\tau} \rho(r,t) dt}{\int_{0}^{\tau} dt}=
\frac{\rho_M (1+e)^3}{\pi}
\int_{\theta_c}^{\pi} \frac{1}{(1-e \ \cos\theta)^2} d\theta= \nonumber \\
 & = &\frac{\rho_M (1+e)^3}{\pi (1-e^2)} [\frac{\pi}{\sqrt{1-e^2}}-
\frac{\sqrt{e^2- \eta^2}}{1-\eta}-\frac{2}{\sqrt{1-e^2}} \tan^{-1}
 (\frac{1+e}{\sqrt{1-e^2}} \sqrt{\frac{e-\eta}{e+\eta}})],
\end{eqnarray}
while between $0$ and $r_c$ (the soft core) the average density is:
\begin{equation} 
\rho_c=\rho_M (\frac{1+e}{1-e})^{3/2}
\end{equation}
where $\eta=1-(1+e)r/r_0$ and $\cos \theta_c=\eta /e$. The 
constant density between $0$ and $r_c$ is product of the
{\it particle angular momentum which prevents the
particles from reaching the center}. The core mass is given by: 
\begin{equation} 
M_c = M(\frac{1-e}{1+e})^{3/2}.
\end{equation}
Equations (5), (11), and (12) clearly show that the ellipticity
of the orbits is at the basis of the soft
core formation in a monolithic collapse. 

From our analysis, we conclude that soft cores in virialized dark 
collisionless halos may result only from two simultaneous conditions: 
(1) an initial homogeneous density profile for the progenitor of 
the present structure (monolithic collapse), and (2) the presence 
of velocity dispersion with a tangential component. The angular 
momentum of kinetic motion avoids the migration of particles toward 
the center, limiting the central density.

In order to obtain a more quantitative relation between the 
core radius \rc\ and the velocity dispersion \vrms, we
resort to N-body simulations of monolithic spherical (top-hat) 
collapse with several values of \vrms\ injected uniformly at 
the maximum of expansion. The public version of HYDRA,
an adaptive P3M-SPH code (Couchman, Thomas \& Pearce 1995),
was used. The simulations can be rescaled to any mass M and 
radius at maximum expansion R$_{\rm M}$ through the following 
relations: M$=m\hat{\rm M}$, R$_{\rm M}=r\hat{\rm R}_{\rm M}$, 
$t=(r^{3}/m)^{0.5} \hat{t}$, and 
\vrms=$(m/r)^{0.5} \hat{\rm v}_{\rm rms}$, where $m$ and $r$ are 
scaling parameters and the quantities with a hat 
are obtained in the simulation. We will assume that 
$\hat{\rm v}_{\rm rms}$ is related to a relict thermal velocity 
$\hat{\rm v}_{\rm th}$, which decays adiabatically as 
$\hat{\rm v}_{\rm th}\propto \hat{\rm R}^{-1}$ until the sphere 
attains its maximum expansion, $\hat{\rm R}=\hat{\rm R}_{\rm M}$. 
We give $\hat{\rm v}_{\rm th}$ when the fluctuation is still in its 
linear regime and calculate its value, $\hat{\rm v}_{\rm th,M}$,
at the maximum of expansion of the sphere of mass $\hat{\rm M}$. 
The latter is the stage from which we start the N-body simulation.

From a set of simulations with $\hat{\rm M}=3.1\times 10^{11}\msun$ 
(1.64 \ $10^4$ particles) and $\hat{\rm R}=0.5$ Mpc, we have 
found how the core radius scales with the injected velocity 
dispersion. The softening radius is 500 pc. We define the core radius 
\rc\ as the radius where the central density decreased by a factor 
of 3. In Fig. 6 we show the results from simulations of hot 
monolithic collapse (open triangles) and the results of the 
analytic model presented above (continuous and dashed line). 
The line is valid for high tangential velocity dispersion when 
shell crossing effects are negligible.
When $\hat{\rm v}_{\rm th}\lesssim 6$ \kms, the linear
relation $\hat{\rc}\approx3$[kpc]$\hat{\rm v}_{\rm th}$ is a good
approximation. Using the scaling relations mentioned above and the 
expression R$_{\rm M}=[(32 GM)(H_0^2 \Omega_m 9 \pi^2)]^{1/3}/(1+z_M)$
valid approximately for a top-hat sphere in a flat low-density universe,
we obtain:
\begin{equation}
\rc\approx\frac{2.5 \kpch}{\sqrt{\Omega_{m,0.3}}} \ 
\frac{{\rm v}_{\rm th,M}/\kms}{(1+z_{\rm M})^{3/2}}.
\end{equation}
The applicability of this formula is valid only for
${\rm v}_{\rm th,M}\lesssim 2$\kms $\Omega_{m,0.3}^{1/6} 
 (hM_{10})^{1/3}(1+z_{\rm M})^{1/2}$, where $M_{10}$ is 
the mass in units of $10^{10}\msun$. 
In the $\Lambda$WDM case, $z_{\rm M}$ refers to the redshift of
maximum expansion of structures with masses close to \mfs, 
and ${\rm v}_{\rm th,M}$ refers to the relict thermal velocity
dispersion these structures have at $z_{\rm M}$.  Note that the
upper limit velocity given above is by much larger than the 
corresponding \vthw. If the less
massive halos, which form by fragmentation, also suffer a  
monolithic collapse at a redshift not much later than
$z_{\rm M}$ (see $\S 6.1$), then their thermal core radii also
can be roughly predicted with eq. (13).  Equation (13) 
is easy understood on the light of the angular momentum analysis
of Bode et al. (2000).

For a $\Lambda$WDM model with \mw=0.6 KeV, then 
$\mfs= 2.8\times10^{11} \msunh$ and \vthw=3.5 \kms\ at
$z=40$. According to the spherical top-hat model
and using the respective $\Lambda$WDM variance, the
maximum expansion redshift for a $\mfs= 2.7\times10^{11} \msunh$
$2\sigma$ ($1\sigma$) fluctuation is $z_{\rm M}=5.3$ (2.1). 
At this redshift the thermal velocity of the top-hat sphere 
decreased adiabatically to ${\rm v}_{\rm th,M}=1$ \kms (0.5 \kms). 
Then, according to eq. (13), the core radius of halos with masses 
close to \mfs\ is $\rc\approx150$ pc (215 pc). This radius should 
be approximately the same for smaller halos, unless these halos 
during the fragmentation process acquire large velocity dispersions. 
Therefore, the thermal soft cores in 
WDM models are only a very small fraction of the virial radius, at 
least for halos not much smaller than \mfs.
                             
\subsection{Results from N-body simulations}

Previous estimates showed that the 
warmon relict velocity dispersion \vthw\ is too small to 
influence the halo inner density profiles. Now we resort to cosmological
simulations which include several values of \vth\ in order to 
explore this question.  
For a warmon of $\sim 1\ \kev$, \vthw\ is small (at $z = 40 $, 
$\vthw \sim 2\ \kms$ which is about 10 times smaller than the typical 
peculiar velocities at that epoch).
A larger thermal velocity dispersion could be possible if warmons  
self-interact; in this case the same free-streaming
scale \fsl\ can be attained with a smaller particle mass (\cite{Hogan99}; 
\cite{Hannestad2000}) and, since $\vthw \propto \mw^{-4/3}$, a smaller
\mw\ implies a larger \vthw. According to Hannestad \& Scherrer (2000),
the mass of the warmons that produce a given free-streaming
scale in the case of collisionless WDM could be 1.9 times smaller 
if warmons self-interact. Therefore, the thermal velocity dispersion 
could be up to 2.3 times larger in the latter case. 

We have run two of the simulations presented in \S 3 (runs \modII\ 
and \modVI) introducing a thermal velocity component in the
particles with several amplitudes.
The thermal velocities were randomly oriented and their 
magnitudes were drawn from a Fermi-Dirac phase space distribution 
with a given rms velocity \vth. These
velocities were added to the initial peculiar velocities
computed using the Zel'dovich approximation. 

For our preferred model (\fsl=0.2 \mpc), we compare the c$_{1/5}$
concentrations obtained in runs \modII\ (\vth=0) and the 
same model but with \vth\ twice higher than $\vthw$,
the warmon velocity corresponding to the given \fsl\ value (\modI), 
making echo of the paper by Hannestad \& Scherrer cited above.
In  Fig. 7 we plot c$_{1/5}$ versus \mvir\ for the only host halo and
the more than a dozen guest halos obtained in both simulations
at two different epochs, $z=0$ and $z=1$. There is not any obvious
difference, in the concentrations of 
halos from both runs. The density profiles of the host and guest 
halos with more than 1000 particles from run \modI\ are shown 
in the lower panel of Fig. 2. As can be seen from Figs. 2 and 7, 
the introduction of a relict velocity dispersion, even two times 
higher than \vthw, does not affect notably the structure  
of WDM halos, at least in the parts where we attain good 
resolution (down to $\sim0.04$ \rv).

In order to explore in more detail the effect of \vth\ on the
inner halo structure, we have run model 
\modVI\ (its resolution is $\sim0.01$ \rv) with an ever increasing 
\vth. In runs \modVIII, \modIX, and \modX, \vth\ was fixed 
to values 2, 4 and 16 times larger than the 
$\vthw$ corresponding to a warmon of mass 125 eV (\fsl=1.7 \mpc), 
respectively. For the latter case, the thermal velocities at 
$z=40$ are $\sim7$ times higher than the peculiar velocities.
Figure 8 shows the density profiles for the four host halos
(shifted vertically by -1 in the log) obtained in the series of runs 
\modVI, \modVIII, \modIX, and \modX. While for the simulations
with \vth= 2 and 4 times $\vthw$ the inner density
profiles still do not deviate significantly from the case
with \vth=0, in the simulation
with $\vth=16\vthw$, a soft core is already evident
at radii smaller than $\sim0.03-0.04$ \rv. 

The redshifts at maximum expansion of the four halos from runs
\modVIII, \modIX, and \modX, are roughly $1.7-1.3$.
Therefore, the core radii predicted with eq. (13)
for these halos are approximately $3.6-3.9$, $7.2-7.8$ 
and $28.6-31.4$ \kpch, respectively. Halos in the last two
runs have roughly $\rc= 10-15$ \kpch\ and $\rc=31-33$ \kpch,
respectively. One should take into 
account that in the cosmological simulations, the studied 
halos do not suffer a perfect monolithic collapse, and that 
some velocity dispersion can be acquired during the collapse 
of the pancakes and filaments.

In conclusion, the introduction of a warmon relict thermal velocity 
has no any important effect on the density profiles 
and concentrations of WDM halos. This velocity would
have to be much larger in order to produce noticeable soft cores, as
eq. (13) shows. We should note that
simulations with and without a thermal velocity are not
similar at all. The identity of the guest halos is not the same
for all of them and their spatial distribution is different in 
both simulations. However, in a statistical sense, neither the 
concentrations nor the satellite circular velocity function
change.  

\section{Discussion}

\subsection{The formation of low-mass halos in simulations with
a damped power spectrum}

When the power spectrum of fluctuations is damped above some
wavenumber $k_f$, ``pancakes'' of size $\sim k_f^{-1}$ are the 
first structures to collapse and smaller objects are expected
to form later by fragmentation (e.g., Zel'dovich 1970; Doroshkevich
et al. 1980). The coupling between different modes in the
non-linear gravitational evolution
drives a transfer of power with power flowing from larger to
smaller scales. This transfer is very efficient;
numerical simulations confirmed previously such a behavior (\eg\ Little,
Weinberg, \& Park 1991; \cite{BP97}; White \& Croft 2000).   

Although our numerical experiments were not aimed at studying 
the structure formation process in a statistical sense (for this
see Bode et al. 2000), we noticed that indeed 
the first structures to collapse are those with scales close 
to the filtering scale length $\lambda_f$. These structures form 
smooth coherent filaments when they enter the non-linear regime 
(Fig. 1a; see also Melott \& Shandarin 1990; Little et al.
1991, and more references therein) instead of the
chains of dense clumps seen in CDM simulations (Fig. 1b).

In the left panel of Fig. 1c one may 
appreciate the filamentary structure of the protohalo that
at $z = 1$ is becoming non-linear. This structure collapses
roughly at the same time (quasi-monolithic collapse) but
obviously the initial conditions are not spherically symmetric. 
Nonetheless, matter flows towards the center of the filament and some
spherical symmetry is established there (see the virialized halo at
$z = 0$). Within the shrinking filament, substructure probably forms 
by fragmentation. Theoretical and numerical support to the idea that 
cosmlogical panckakes are unstable with respect to fragmentation
and the formation of filaments can be found in Valinia et al. (1997),
and more references therein.

Based on a visual inspection (see also Table 2 and discussion below), 
we may say that the collapse of the substructure within the filament
is almost parallel to the collapse of the filament. 
As mentioned above, the power transfer from larger to smaller scales 
is very efficient. It also seems that the collapse epoch of the
fragmented halos is independent of their masses. Nevertheless,
all these halos collapse on average later than in a CDM simulation.
This might explain why the small WDM halos are less concentrated
and why their concentrations, though with a large scatter, do not
depend on mass (see Fig. 5). Because halos are less concentrated
than in the CDM simulations, they are more easily disrupted. As was
shown in \Colinetal, the reduction of the number of small halos due
to this effect is comparable to the effect that the power spectrum
suppression has on this number, and both work together to 
deliver a very small number of satellites at $z = 0$. 

Interestingly, our results suggest that on average and in a first 
approximation, the density profiles of dark matter halos are indeed 
universal, no matter how they form, either by hierarchical clustering 
or by a monolithic collapse or by fragmentation (see also Moore et al.
1999b). The difference seen in their 
concentrations can be explained just as an {\it effect of the formation 
epoch, which for small WDM halos is delayed compared to the
formation epoch of their CDM counterparts}. As we have shown in 
\S 5, a way to affect significantly the inner density profile of halos
in the case of a monolithic collapse is by including a high (tangential)
velocity dispersion. The thermal velocities of warmons of mass 
$\sim 1\ \kev$ are much smaller than this required velocity dispersion.

\subsection{Viability of the WDM scenario: observational tests}

Because of the increasing evidence that the predictions of the nowadays
standard $\Lambda$CDM model at small scales are in conflict with observations,
modifications to this scenario, {\it able to retain their successful
predictions at large scales,} have been recently analyzed. As the nature
of dark matter particles is still a mystery, it is tempting to
exchange CDM particles for WDM particles as 
the most simple modification 
to the standard scenario. At least from the point of view of the particle 
physics, there is not an obvious preference for any of 
these particles (\eg\ \cite{CDW96}). But,
what are the advantages of the WDM scenario with respect to the 
CDM one from the point of view of structure formation? Following,
we present a list of what we consider are these advantages:

1. For a WDM model with $\fsl \simgreat 0.1\ \mpc$ ($\mw \lesssim 1$ KeV), 
$\Omega_m = 0.3$, and $\Omega_\Lambda = h = 0.7$, the observed maximum circular 
velocity function of Milky Way and Andromeda satellites is roughly 
reproduced (\Colinetal).

2. Although in the WDM scenario the halos with masses close to or
below \mfs\ do not have a noticeable constant density core, they 
are less concentrated and have a density profile 
shallower in the center than their CDM counterparts (\S\S 4 and 5). 
Recent observational studies have shown that with the current 
data is not possible to accurately constrain  the halo inner 
density profiles of dwarf and LSB galaxies; these profiles are probably 
not steeper than $r^{-1}$ and, if anything, the concentrations of these 
halos are lower than those predicted in the CDM scenario (van den Bosch
et al. 2000; Swaters, Madore \& Trewhella 2000; van den Bosch
\& Swaters 2000). In Fig. 9 we compare the c$_{1/5}$ concentration
of guest WDM halos from runs \modI\ and \modII\ with the c$_{1/5}$
concentration of dwarf and LSB galaxies, inferred from
observational data (see details in the figure caption). Unfortunately,
there is not an overlap between the theoretical and observational data; 
nonetheless, it can already be appreciated that the guest
WDM halos are in better agreement with observations.

3. The formation of disks within WDM halos ($\mw \lesssim 1\ \kev$)
in N-body+hydrodynamic simulations does not seem to suffer from the disk
angular momentum problem (\cite{SomDol00}). This problem is also at the
basis of other difficulty reported by Steinmetz \& Navarro (1998): the
predicted infrared Tully-Fisher (TF) relation in their simulations is 
much brighter than the observed one. The maximum circular velocity
$V_{\rm max}$ of the system increases after disk formation. In the 
numerical simulations of Steinmetz \& Navarro this increase is about
a factor of two larger than for models where detailed angular momentum
conservation is assumed for the infalling gas (e.g., Mo, Mao, \& White
1998; Avila-Reese et al. 1998; Avila-Reese \& Firmani 2000; Firmani 
\& Avila-Reese 2000). A factor of two in the velocity translates into
a factor of $\sim 8$ in mass or luminosity explaining why Steinmetz
\& Navarro obtain a brighter TF relation. If the angular momentum
problem is alleviated as in the WDM scenario, then one expects that the 
zero-point of the TF relation predicted in numerical simulations
will be in agreement with observations (\cite{SomDol00}). 
  

4. Several authors have shown that the TF relation 
of normal disk galaxies in the infrared band is an imprint of 
the mass-velocity ($\mvir-V_{\rm max}$) relation of CDM halos 
(Firmani \& Avila-Reese 2000 and more references therein). For 
masses larger than $\sim 10 \mfs$, there are not major 
differences between the CDM and the WDM halos (\Colinetal; 
see also Avila-Reese et al. 1998). How does the $\mvir-V_{\rm max}$ 
relation look for small masses? 
In Fig. 10 we have plotted $\mvir$ versus $V_{\rm max}$ 
for our WDM models with $\fsl = 0.2\ \mpc$. We also show in this
figure the $\mvir - V_{\rm max}$ relation of guest halos obtained 
for a $\Lambda$CDM simulation, its scatter and a linear fit to 
this relation (Avila-Reese et al. 1999). As expected, the WDM 
halos move towards the lower $V_{\rm max}$ side. The formation 
of dwarf galaxies may be strongly affected by feedback and 
reionization; these galaxies could lose some of their initial 
baryon matter through galactic winds (\eg\ Mac Low \& Ferrara 1999) 
or through photo-evaporation (Shapiro \& Raga 2000) and thus see 
their luminosities diminished. In the TF plot this means that for 
a given luminosity, $V_{\rm max}$ shifts to larger values. 
Observations show that dwarf and normal galaxies have roughly the 
same infrared TF relation; if anything, dwarf galaxies lie more on 
the low-velocity side and present more scatter
(Pierini \& Tuffs 1999; de Jong \& Lacey 2000, quoted by
\cite{CLBF00}, see their Fig. 7). Therefore, the fact that for 
WDM the small halos are shifted in the $\mvir - V_{\rm max}$ 
relation as it is shown in Fig. 10 might resolve a potential 
problem of the CDM scenario. Therefore, it seems that the 
{\it WDM scenario could reproduce the infrared TF relation of dwarf 
galaxies better than the CDM scenarios does}.

5. Unlike in the CDM scenario, for WDM one expects that the 
satellite dwarf galaxies form later than the host large galaxies. The 
evolution of the substructure for our WDM simulation 
\modI\ (\fsl=0.2 Mpc) is shown in Table 2. The scale factor normalized to
one at present is given in the first column while the number 
of halos with maximum circular velocity $V_{\rm max}$ greater than 
100 \kms\ is presented in column 2. Columns 4, 5, and 6 give the 
coordinates of the center of mass of the system composed by these halos. 
The total number of halos 
with $V_{\rm max} > 15\ \kms$ that contain more than 200 particles and 
the mass of the most massive halo are given in columns 3 and 7, 
respectively. We see that the first structure starts to assemble at 
$z \sim 7$; this seed has a mass of $\sim 3 \times 10^{10}\ \msunh$.
Later, at $z \sim 4$, when the mass of the most massive halo 
has grown to $\mvir\sim10^{11}\ \msunh$, 
the first substructures appear by 
fragmentation. Thus, the building blocks for large galactic and 
supra-galactic structures are those with masses near to \mfs; smaller
structures (guest halos) form slightly later by fragmentation. 
We may say thus that the formation redshift of guest halos
is approximately equal or less than $z_f (\mfs) $. For the 
$\Lambda$WDM model with the power spectrum normalized to 
COBE used here and according to the 
spherical top-hat model, the typical formation redshift of $\sim10^{11}\ 
\msunh$ halos is $z_f(10^{11}\msunh)\approx 1.3$ and 3.7 for 
$1-\sigma$ and $2-\sigma$ peaks, respectively. If galaxies 
form from high peaks, then the typical formation redshifts of galaxies 
of $\sim10^{11} \msunh$ will be larger than $z=2-3$. For the $\Lambda$WDM
models with $\fsl \lesssim 0.2\ \mpc$, smaller galaxies (dwarfs) will form 
slightly later than these redshifts. Similar conclusions are obtained
by Bode et al. (2000), who also remark that low mass halos form solely
within pancakes and filaments, and not in the voids, in contrast
to the situation in the CDM scenario (see also Fig. 1).

From the observational point of view, there are
some pieces of evidence that dwarf galaxies formed later than 
bright galaxies. In the Local Group, all dwarf galaxies seem to have their
oldest stellar population slightly younger than the oldest Milky
Way halo population (Mateo 1998); for the Large Magellanic Cloud this
age difference could be of 2 Gyr, according to studies of the horizontal
branch morphology of stellar clusters (Olszewski, Suntzeff, \& Mateo
1996, and references therein), while for the Small Magellanic Cloud
the difference should be even larger since this galaxy seems to be younger
than its neighborhood. On the other hand, comparisons of observed
and modeled galaxy counts for $B$ dropout galaxies at $3.5\lesssim z\lesssim
4.5$ suggest that some $L_{\star}$ galaxies were already in place
at $z\approx 4$ but dwarf galaxies may have formed later at 
$3\lesssim z\lesssim 4$ (Metcalfe et al. 2000), in agreement with 
predictions of the WDM scenario.
  
6. The Lyman$-\alpha$ forest is a powerful probe of 
the linear power spectrum on galactic and sub-galactic scales, upon the 
understanding that it traces the underlying matter density. Since in a 
WDM scenario the small-scales modes are damped out, one might 
think that the observed Lyman$-\alpha$ forest should not be
reproduced by this cosmology. However, because of the efficient 
power transfer from larger to smaller scales (see $\S 6.1$), 
enough power on small scales is regenerated at $z \approx 3-4$.
\cite{NSDCh2000} have shown that WDM models ($\Omega_m=0.3$
and  $\Omega_\Lambda = h =0.7$) with $\fsl \lesssim 0.155\ \mpc$ 
($\mw \simgreat 0.75\ \kev$) are able to reproduce the observed 
properties of the Lyman$-\alpha$ forest. This last point, more
than an advantage of WDM with respect to CDM, is a test for the
former scenario. The CDM scenario is also able to predict the 
properties of the Lyman$-\alpha$ forest.

 As we have seen, the WDM scenario works better than
the CDM one in several aspects. Nonetheless, both scenarios
fail in predicting soft halo cores and the independence of the halo
central density on its mass, as some inferences from observations 
seem to suggest (e.g., Firmani et al. 2000a,b and the references
therein). More observational data which explore in detail the 
inner structure of the halo component of galaxies are urgent\footnote{
At the time this paper has being refereed, new high resolution rotation
curve observations for LSB galaxies were presented by de Blok et al. 
(2001); they conclude that observations are consistent with constant
density halo cores of ''modest ($\sim 1$ kpc) core radius, which
can give the illusion of steep cusps when insufficiently resolved''.}; 
however, it is likely that the more decisive data will 
come from strong lensing studies of clusters of galaxies
(the largest virialized structures). 
The phase space density of dark matter halos derived from
observations also offers an important test for the WDM scenario
(Hogan \& Dalcanton 2000; Sellwood 2000). If more detailed
observational studies confirm the existence of soft halo cores
with the scaling properties suggested in Firmani et al. 
(2000a,b) and Sellwood (2000), the WDM scenario
must be abandoned. Then alternatives such as the
self-interacting CDM might become more appealing.      
 
\section{Summary  and conclusions}

We have carried out high-resolution N-body cosmological simulations
with the aim to study the effect on the halo density 
profiles produced by (i) the damping of the power spectrum 
at small scales and (ii) the introduction of a relict thermal 
velocity dispersion, as well as to explore the viability
of the WDM scenario. We have also studied hot monolithic 
collapse by means of an analytical model and top-hat N-body 
simulations. Our main conclusions are:

-For the halos studied here, with masses close or up to $\sim100$ 
times smaller than the filtering mass \mfs, the density profiles 
are on average well described by the NFW shape; the density
profiles were well resolved down to $\sim0.01$ \rv.
The only differences between halos with masses below \mfs\ and their 
CDM counterparts is that the former have lower concentrations and
their innermost density profiles are not as steep as $r^{-1.5}$. Thus, 
though the cosmogony of the halos in both cases is different, the 
final virialized structures are not dramatically different. 

-The c$_{1/5}$ or c$_{\rm NFW}$ concentrations of halos more 
massive than a few times the filtering mass \mfs\ are similar to 
those of their CDM counterparts. However, as the mass decreases, 
the concentrations on average remain almost constant (slightly 
decrease), while for CDM halos the concentration increase 
monotonically. The scatter of the concentration for a given mass 
is large in both cases. The difference in the concentrations
of the small halos can be explained because the 
relatively late formation epoch of small halos in the  
simulations with damped power spectrum at small scales.
   
-The relict thermal velocity dispersion of warmons, \vthw,
does not affect the density profiles of WDM halos, at 
least down to 0.01 \rv. For our simulations with \fsl=0.2 and 
1.7 Mpc (\modI\ and \modVI, respectively) we used a thermal 
velocity two times larger than \vthw\ and we did not find any 
significant difference in the halo concentrations and density 
profiles with respect to the case with \vth=0. For the 
high-resolution run \modVI\ we also experimented with 
\vth= 4 and 16 times \vthw, finding resolved soft cores only 
for the last case (with radii at $\sim0.03-0.04$ \rv).

-The inner structure of dark matter halos formed through monolithic
collapse can be affected {\it only} if the particles have a significant
(tangential) velocity dispersion \vrms. The penetration degree towards
the halo center is determined by the pericenter of the orbiting particle,
or equivalently, the angular momentum forces the particles to avoid 
migration towards inner regions. From N-body simulations
of the collapse of top-hat spheres with different amounts of \vth,
we have found that the core radius scales linearly with \vth\
up to velocities much larger than \vthw. In order that
WDM halos with masses close to \mfs\ form noticeable cores,
\vth\ should be much larger than \vthw. 

We conclude that a flat $\Lambda$WDM model with $\fsl\approx 0.16$ Mpc 
($\mw\approx 750$ eV), $\Omega_{\Lambda}\approx0.7$, and $h\approx 0.7$,
has several advantages over its CDM model counterpart.
The problem of excess of substructure is solved (\Colinetal),
the halos of dwarf and LSB galaxies are less concentrated
than in the CDM scenario as observations suggest (Fig. 9), and the 
disk angular momentum problem is alleviated (\cite{SomDol00}). 
Furthermore, our preferred WDM model describes
better than CDM the TF relation of dwarf galaxies 
(Fig. 10) and the formation epochs of these galaxies (\S 6),
as well as their large-scale distribution (Bode et al. 2000).
On the other hand, this model is not apparently
in conflict with measurements of the power spectrum
of Lyman$-\alpha$ forest (Narayanan et al. 2000) and with
the reionization constraints (Bode et al. 2000). 
Nonetheless, the WDM model is ruled out if
more observations confirm the existence of soft halo cores as
well as the independence on mass of their densities; the crucial test 
is at cluster scales. No doubt, exciting questions
remain to be answered in the near future.

\acknowledgments
 
We are grateful to A. Klypin and A. Kravtsov for kindly 
providing us a copy of the ART code in its version of multiple mass,
and to H. Couchman for having made available his adaptive P3M-SPH 
code HYDRA. We also thank A. Klypin for enlightening discussions.
We thank the second referee for a throughout and accurate 
revision of the manuscript as well as for the suggestions which 
helped to improve the quality of this paper. The first referee is 
acknowledged for his (her) criticism that conducted us to carry out more simulations
in order to study the influence of thermal velocity on halo density profiles.
This work has received partial funding from CONACyT grant J33776-E to V.A. 
The work of E.D. was supported by Fondazione Cariplo, Italy. 
Our ART simulations were performed at the Direcci\'on General de 
Servicios de C\'omputo Acad\'emico, UNAM, using an Origin-2000 computer. 

\newpage


\newpage

\begin{deluxetable}{cccccrcc}
\tablecolumns{7}
\tablewidth{0pc}
\tablecaption{Models and Simulations Parameters}
\tablehead{\colhead{Run} & \colhead{\fsl} & \colhead{\mfs} & \colhead{\vth
\tablenotemark{a}} & 
\colhead{$m_p$} & \colhead{Box} & \colhead{Resolution} & \colhead{$N_{host}$} \\
   & (\mpc)  & \msunh & ($\times\vthw$) & (\msunh) & (\mpch) & (\kpch) &  }
\startdata
\modII & 0.2 & $2.9 \times 10^{11}$ & 0.0 & $2.1 \times 10^6$ & 7.5\phm{100}  & 0.1 & 1 \nl
\modIII & 1.7 & $1.7 \times 10^{14}$ &  0.0 & $2.1 \times 10^6$ & 7.5\phm{100}  & 0.4 & 1 \nl
\modIV & 1.7 & $1.7 \times 10^{14}$ & 0.0 & $1.7 \times 10^7$ & 15.0\phm{100} & 0.4 & 1 \nl
\modV & 1.7 & $1.7 \times 10^{14}$ & 0.0 & $1.3 \times 10^8$ & 30.0\phm{100} & 1.8 & 4 \nl
\modVI & 1.7 & $1.7 \times 10^{14}$ & 0.0 & $1.1 \times 10^9$ & 60.0\phm{100} & 1.8 & 4 \nl
\modVII & 0.0 & 0.0 & 0.0 & $1.1 \times 10^9$ & 60.0\phm{100} & 1.8 & 4 \nl
\modI & 0.2 & $2.9 \times 10^{11}$ & 2 & $2.1 \times 10^6$ & 7.5\phm{100}  & 0.1 & 1 \nl
\modVIII & 1.7 & $1.7 \times 10^{14}$ & 2 & $1.1 \times 10^9$ & 60.0\phm{100} & 1.8 & 4 \nl
\modIX & 1.7 & $1.7 \times 10^{14}$ & 4 & $1.1 \times 10^9$ & 60.0\phm{100} & 1.8 & 4 \nl
\modX & 1.7 & $1.7 \times 10^{14}$ & 16 & $1.1 \times 10^9$ & 60.0\phm{100} & 1.8 & 4 \nl
\enddata
\tablenotetext{a} {Thermal velocity dispersion added to the particles at $z=40$,
in units of the relict thermal velocity corresponding to the warmon
mass used in the simulation, \vthw} 

\end{deluxetable}

\begin{deluxetable}{ccrcccc}
\tablecolumns{7}
\tablewidth{0pc}
\tablecaption{Subestructure evolution in model \modI}
\tablehead{\colhead{Epoch} & \colhead{$N_{h,100}$} & \colhead{$N_N$} & 
\colhead{$X_{cm}$} & \colhead{$Y_{cm}$} & \colhead{$Z_{cm}$} & \colhead{$\mvir$} \\
 &  &  & (\mpch) & (\mpch) & (\mpch) & (\msunh) }
\startdata
1.000 &   1 &  13\phm{100} &   7.193 &   5.236  &  1.406 &  $1.4 \times 10^{12}$ \nl
0.879 &   1 &  11\phm{100} &   7.191 &   5.287  &  1.424 &  $1.3 \times 10^{12}$ \nl
0.753 &   1 &  17\phm{100} &   7.175 &   5.349  &  1.443 &  $1.1 \times 10^{12}$ \nl
0.630 &   2 &  24\phm{100} &   7.153 &   5.453  &  1.452 &  $6.6 \times 10^{11}$ \nl
0.504 &   4 &  24\phm{100} &   7.191 &   5.454  &  1.626 &  $5.2 \times 10^{11}$ \nl
0.405 &   4 &  29\phm{100} &   7.139 &   5.705  &  1.516 &  $4.9 \times 10^{11}$ \nl
0.354 &   5 &  26\phm{100} &   7.155 &   5.740  &  1.528 &  $2.9 \times 10^{11}$ \nl
0.303 &   4 &  20\phm{100} &   7.170 &   5.875  &  1.472 &  $2.4 \times 10^{11}$ \nl
0.255 &   3 &  15\phm{100} &   0.379 &   5.837  &  1.515 &  $1.8 \times 10^{11}$ \nl
0.204 &   2 &   6\phm{100} &   7.117 &   6.623  &  1.055 &  $1.1 \times 10^{11}$ \nl
0.153 &   2 &   2\phm{100} &   7.135 &   6.148  &  1.406 &  $5.2 \times 10^{10}$ \nl
0.129 &   1 &   1\phm{100} &   7.151 &   6.994  &  0.650 &  $2.7 \times 10^{10}$ \nl
\enddata
\end{deluxetable}

\newpage
\clearpage 

\begin{figure}
\figurenum{1(a)}
\figcaption{ Distribution of dark matter particles in run 
\modVI\ inside a cube of 40 \mpch\ comoving on a side at different 
epochs: $a=0.3$, 0.5, 0.75, and 1.0, where $a$ is the scale factor. All 
particles that are within a sphere of radius 1.5 \mpch\ centered on a host 
halo at $a = 1$ were traced back to the chosen epoch and their center of
mass was computed. Cubes were then centered on these centers of masses.
We have color-coded particles on a gray scale according to the logarithm 
of their local density (a {\it pgplot} program kindly provided by A. Kravtsov). 
The local density at the particle positions, on the other hand, was computed
using SMOOTH, a publicly available code developed by the HPCC group in the 
University of Washington Department of Astronomy.}
\end{figure}

\begin{figure}
\figurenum{1(b)}
\figcaption{Same as {\it (a)}, but for CDM, run \modVII.}
\end{figure}

\begin{figure}
\figcaption{(c). {\it  Right panel:} Distribution of dark matter particles 
inside a sphere of radius $1.5\ \mpch$ centered on the large host halo 
seen in plot 1(a) at $a = 1$ . The mass of the selected host halo is about 
half the corresponding filtering mass ($\mfs = 1.7 \times 10^{14}\ \msunh$), 
and it was one of the first structures to collapse in the simulation. 
{\it Left panel: }Same particles of right panel but traced back to $a = 0.5$.
All of them lie within a sphere of radius 5.6 \mpch\ comoving centered on 
the center of mass of the particle system at this epoch.}
\end{figure}

\begin{figure}
\plotone{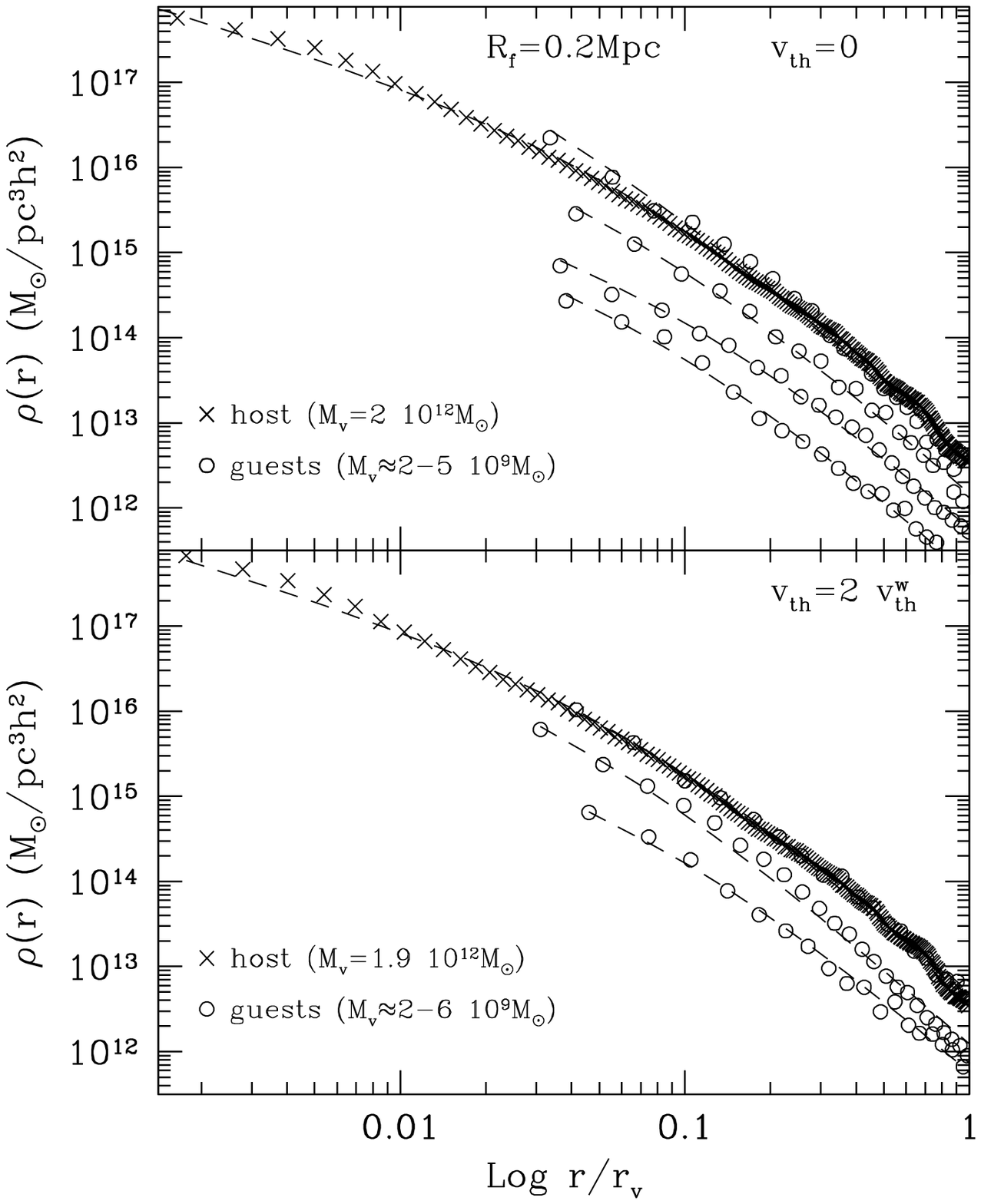}
\figcaption{Density profiles of the host (crosses) and guest halos
with more than 1000 particles (circles) from $\Lambda$WDM simulations 
with \fsl=0.2 \mpc\ (\mw=605 eV) with zero thermal velocity
dispersion (upper panel, run \modII) and with a thermal velocity
dispersion twice larger than that corresponding to a warmon of 605
eV (lower panel, run \modI). Radii are normalized to the
current halo virial radius \rv. The inner point
in each plotted profile is the maximum of 4 times the 
formal force resolution (Table 1) and the radius of the first
point containing more than 50 particles.}
\end{figure}

\begin{figure}
\plotone{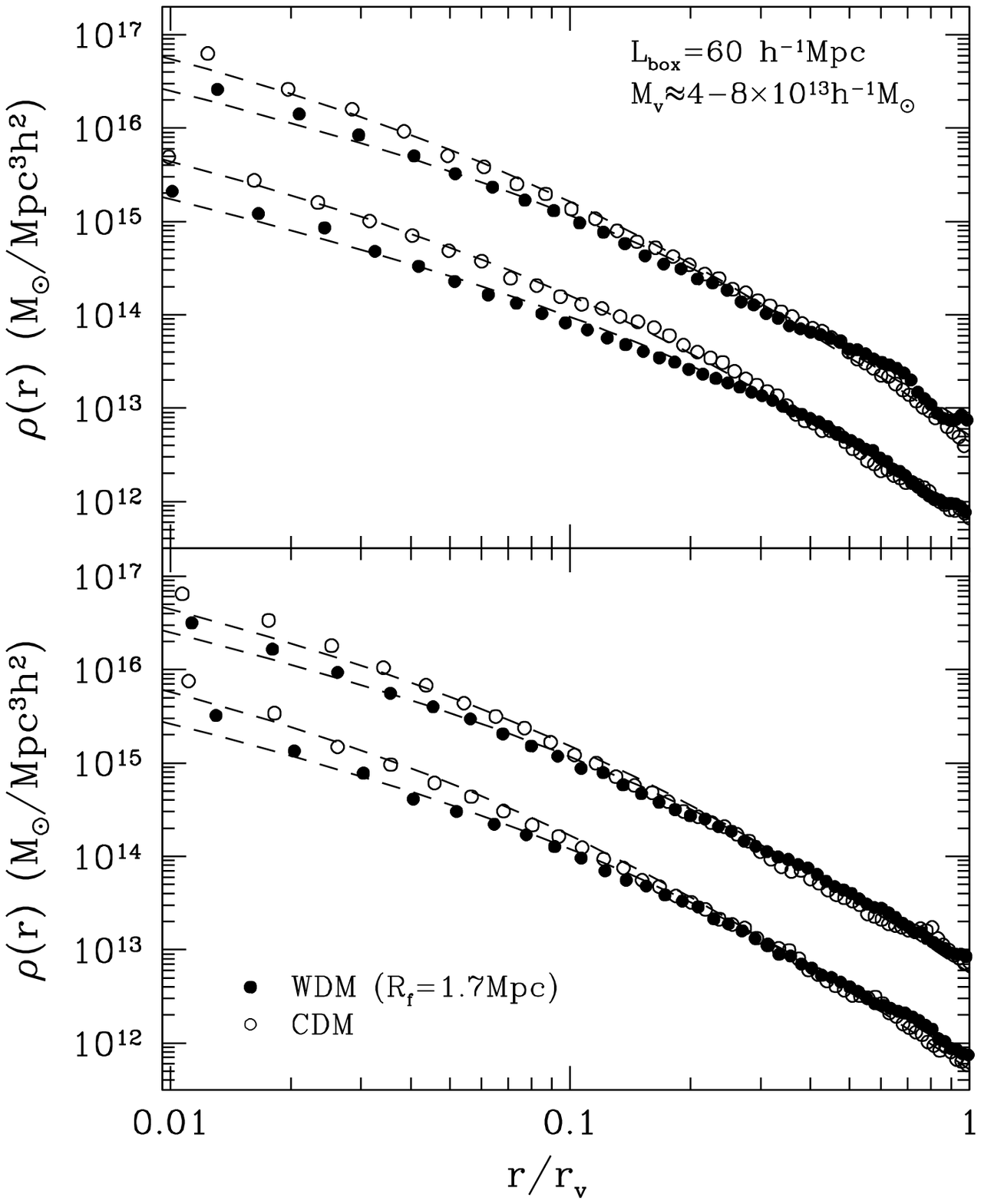}
\figcaption{Density profiles of the four host halos in simulations 
\modVI\ (solid circles) and \modVII\ (empty circles). In order to prevent 
visual confusion, two panels are used and the two lower profiles in each 
0panel were shifted by -1 in the log. The
dashed lines are NFW fits to these profiles.  The inner points and
the normalization of the radius are as in Fig. 2. The masses of 
the WDM halos are 2-4 times smaller
than the filtering mass $\mfs=1.7\times 10^{14}\msunh$. }
\end{figure}

\begin{figure}
\plotone{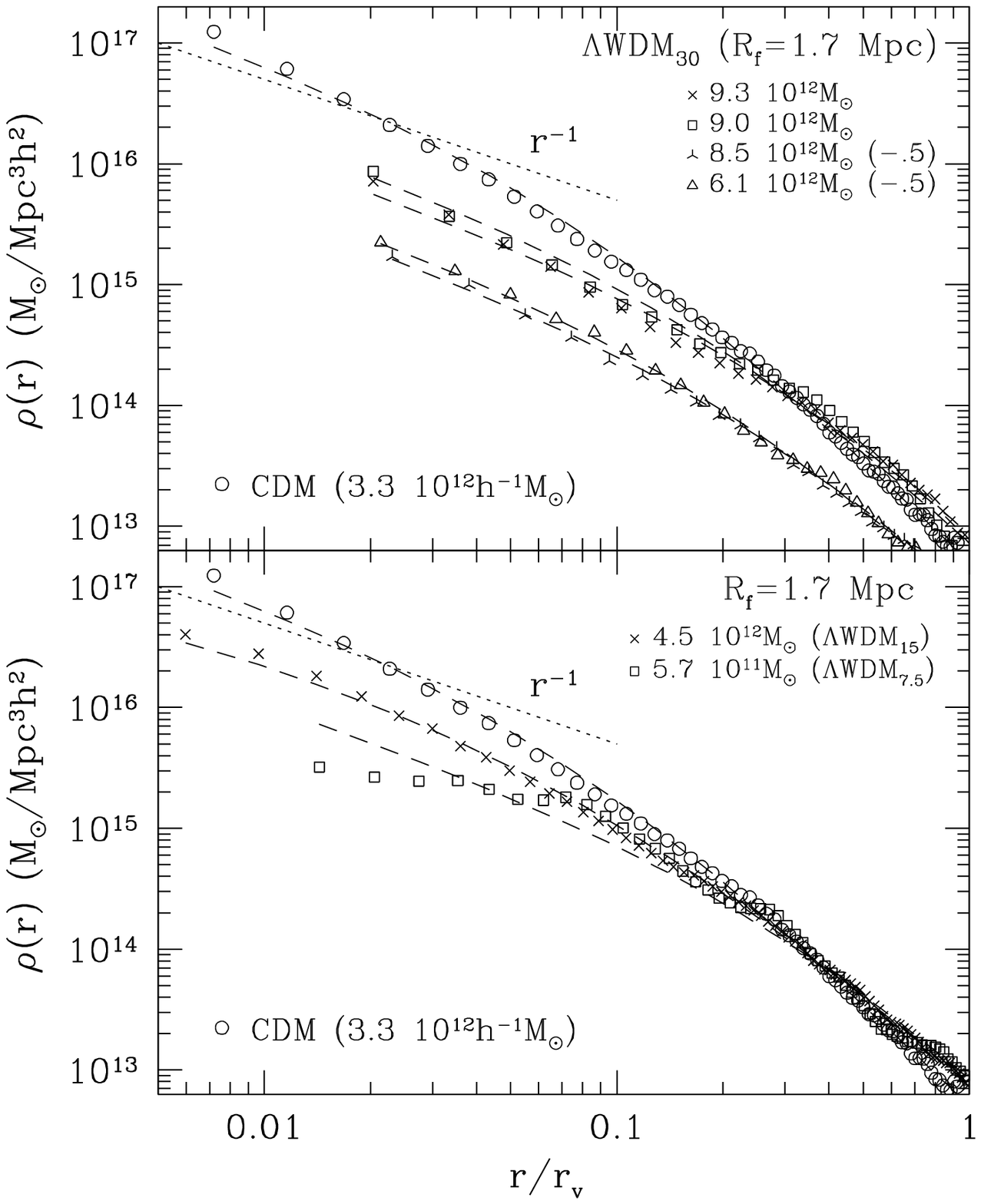}
\figcaption{{\it Upper panel:} Density profiles of the four host 
halos from simulation \modV. The two lower profiles were shifted
by $-0.5$ in the log in order to avoid overlapping. Dashed
lines are the best NFW fit. For comparison, the profile of a 
CDM halo from \Colinetal\ is also plotted (empty circles). The 
inner points and the normalization of the radius are as in Fig. 2.
{\it Lower panel:} The same as the upper panel
but for the two host halos from runs \modIV\ (crosses) and
\modIII\ (squares), respectively. As one expects, the density
profile of the halo from run \modIII\ is affected
by the small size of the box (see \S 3).}
\end{figure}

\begin{figure}
\vspace{17cm}
\includegraphics{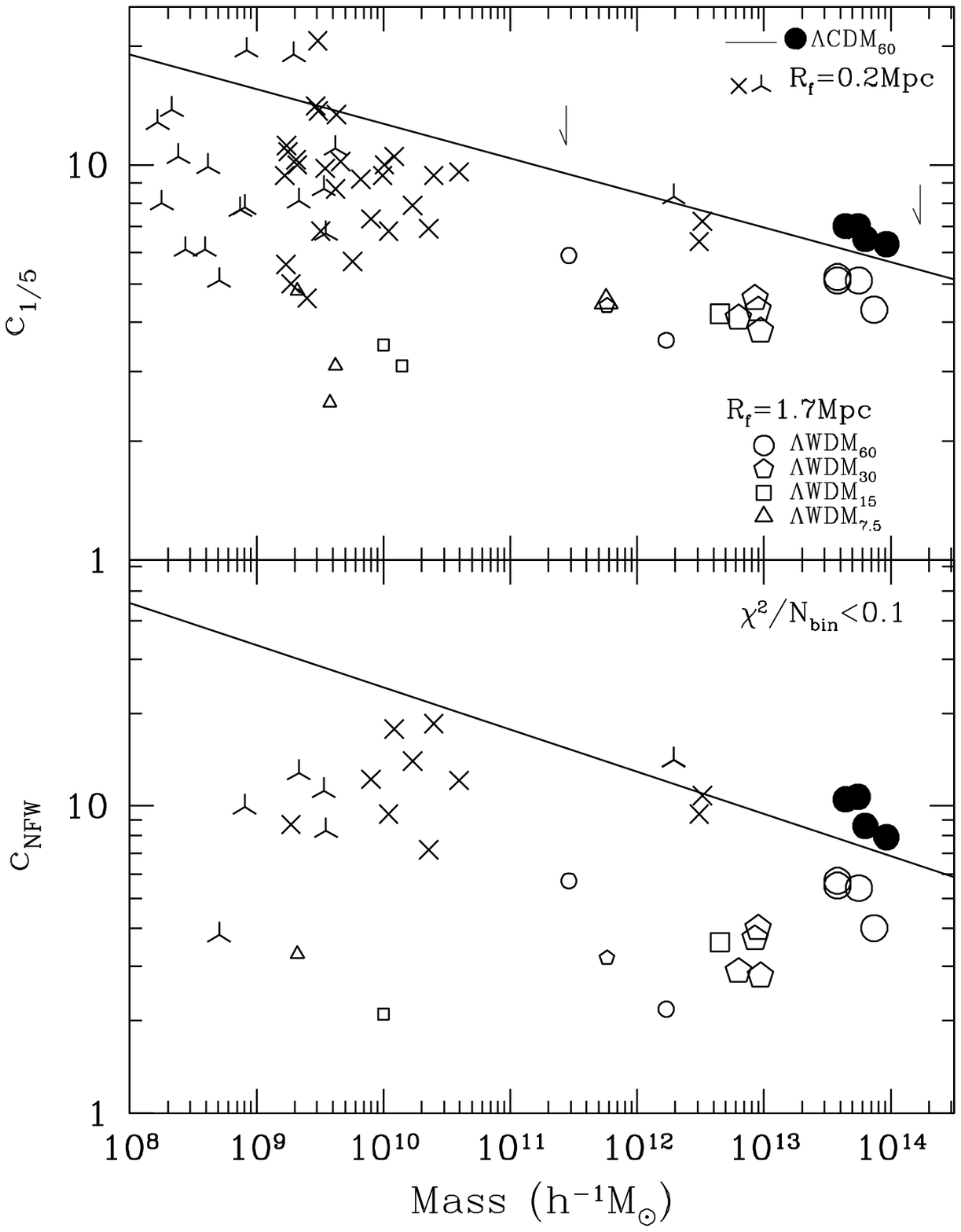}
\figcaption{Concentration parameters c$_{1/5}$ and c$_{\rm NFW}$ 
vs. halo virial mass ({\it upper} and {\it lower} panels respectively) 
from the $\Lambda$WDM simulations with \fsl=1.7 Mpc (empty symbols) and 
\fsl=0.2 Mpc (crosses and skeletal triangles, from \Colinetal, and from 
the model \modII\ presented here, respectively), and from the \modVII\ 
simulation (solid circles). Large empty symbols are for the corresponding 
host halos, while small empty symbols are for the guest halos with more 
than 1000 particles; they are only a few. The solid line is a linear
fit for thousands of isolated halos obtained in a $\Lambda$CDM 
simulation (Avila-Reese et al. 1999). The vertical arrows indicate
the filtering masses \mfs\ corresponding to \fsl=0.2 and 1.7 Mpc.
The c$_{\rm NFW}$ parameter for the guest halos in the simulations of 
\Colinetal\ and the model \modI\ presented here are shown only for those 
halos whose profiles fit the NFW shape with an accuracy better than
$D < 10$\%. Note that the concentration parameters of halos below
\mfs\ remain almost constant as the mass decreases while in the case
of the CDM model these concentrations continuously increase as the
mass decreases.}
\end{figure}

\begin{figure}
\plotone{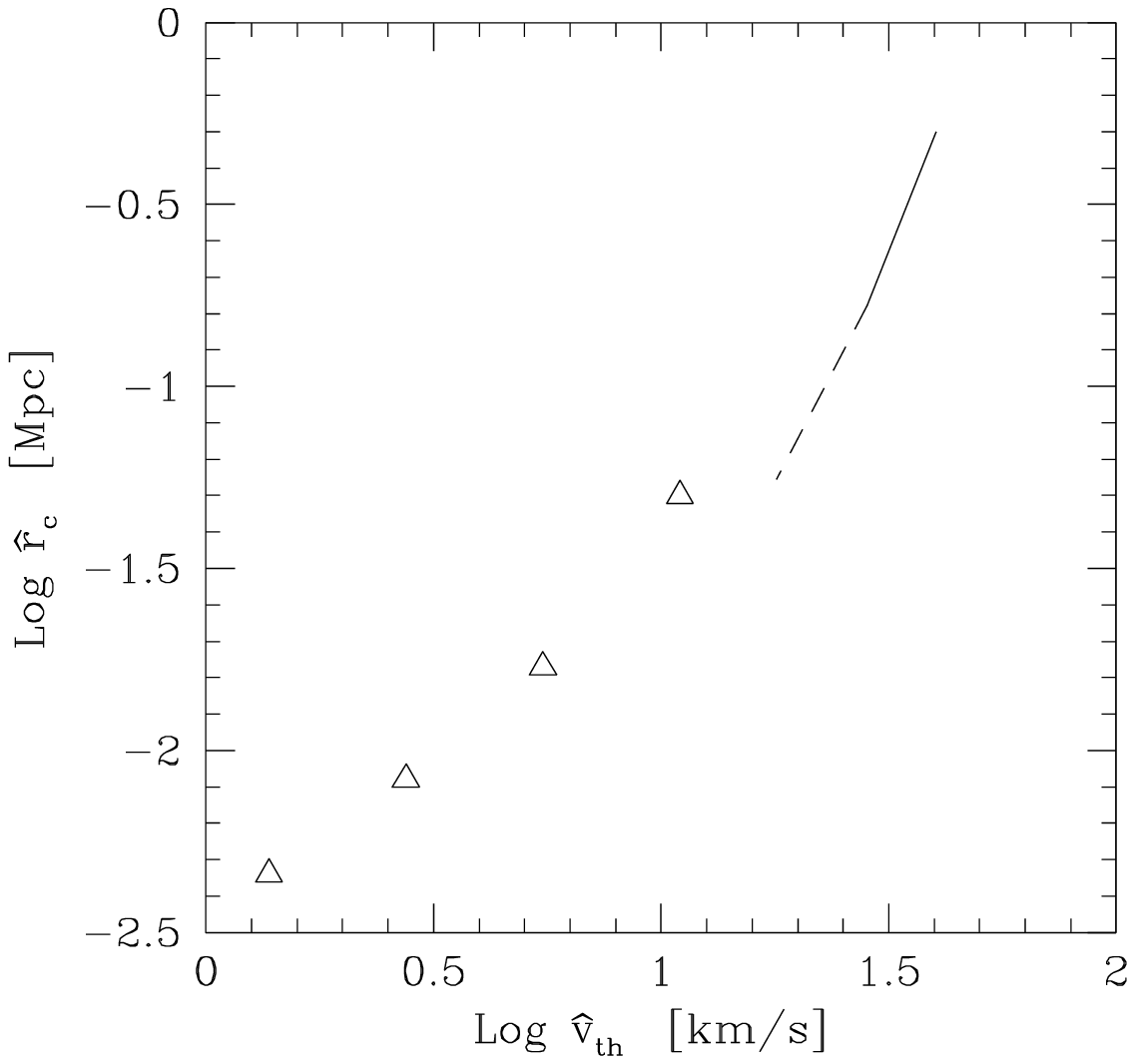}
\figcaption{Core radii obtained in N-body simulations of monolithic
collapse with increasing values of thermal velocity (triangles) and
predicted by the analytical model (line, see text). The dashed part
of the line is at values of thermal velocity where the analytical 
model begins to lose its validity.} 
\end{figure}

\begin{figure}
\plotone{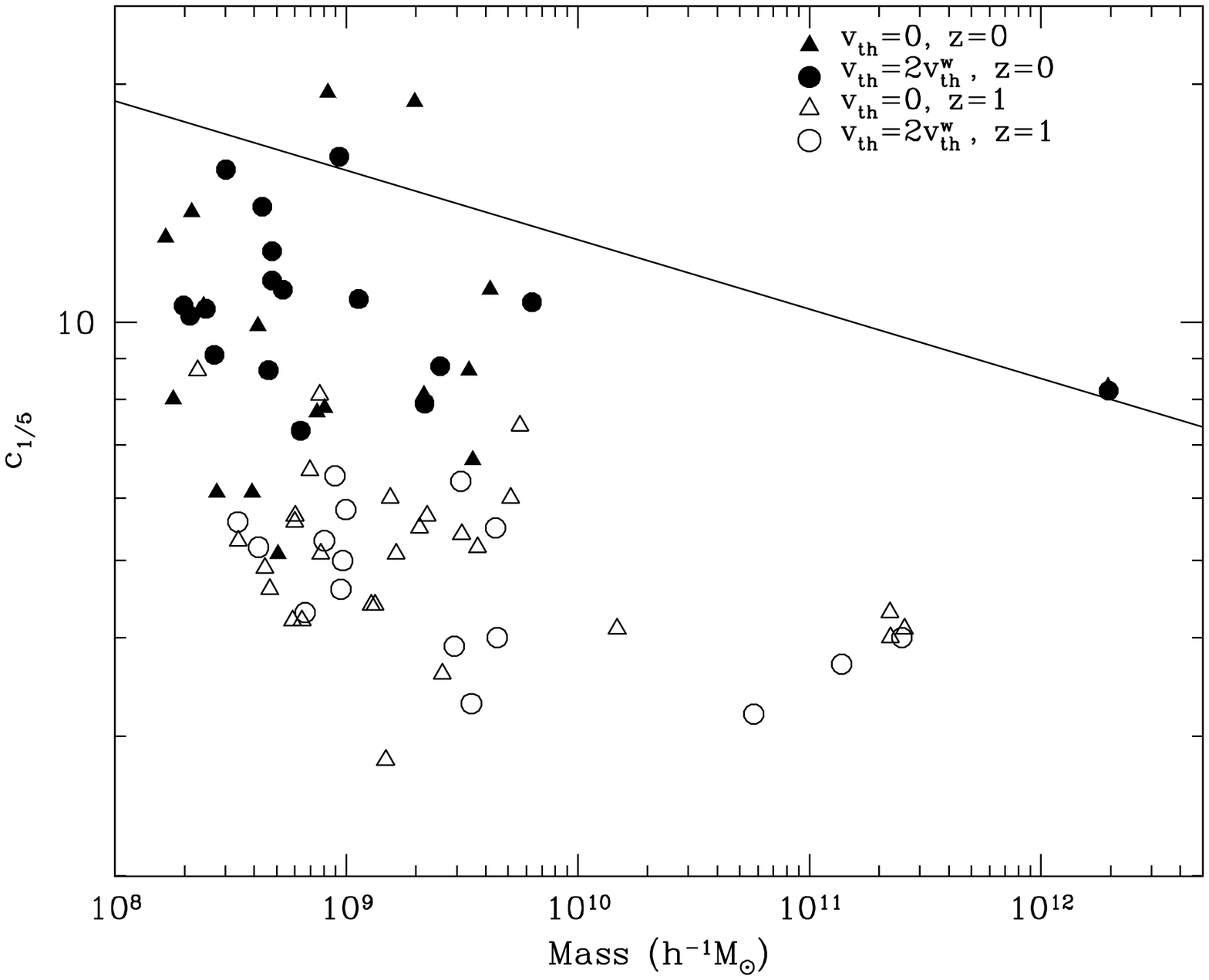}
\figcaption{Concentration parameter c$_{1/5}$  vs. halo mass
from $\Lambda$WDM simulations (\fsl=0.2 Mpc, $L_{\rm box}=7.5 \mpch$)
with \vth=0 (run \modII\, triangles) and with \vth=6.7 \kms determined
at $z_i=40$ (run \modI\, circles); this velocity is two times 
larger than the corresponding \vthw. The filled symbols are for 
halos at $z=0$ while the empty symbols are for halos at $z=1$. There
is not any significant difference in between simulations with and 
without thermal velocity inclusion.}
\end{figure}

\begin{figure}
\vspace{18cm}
\includegraphics{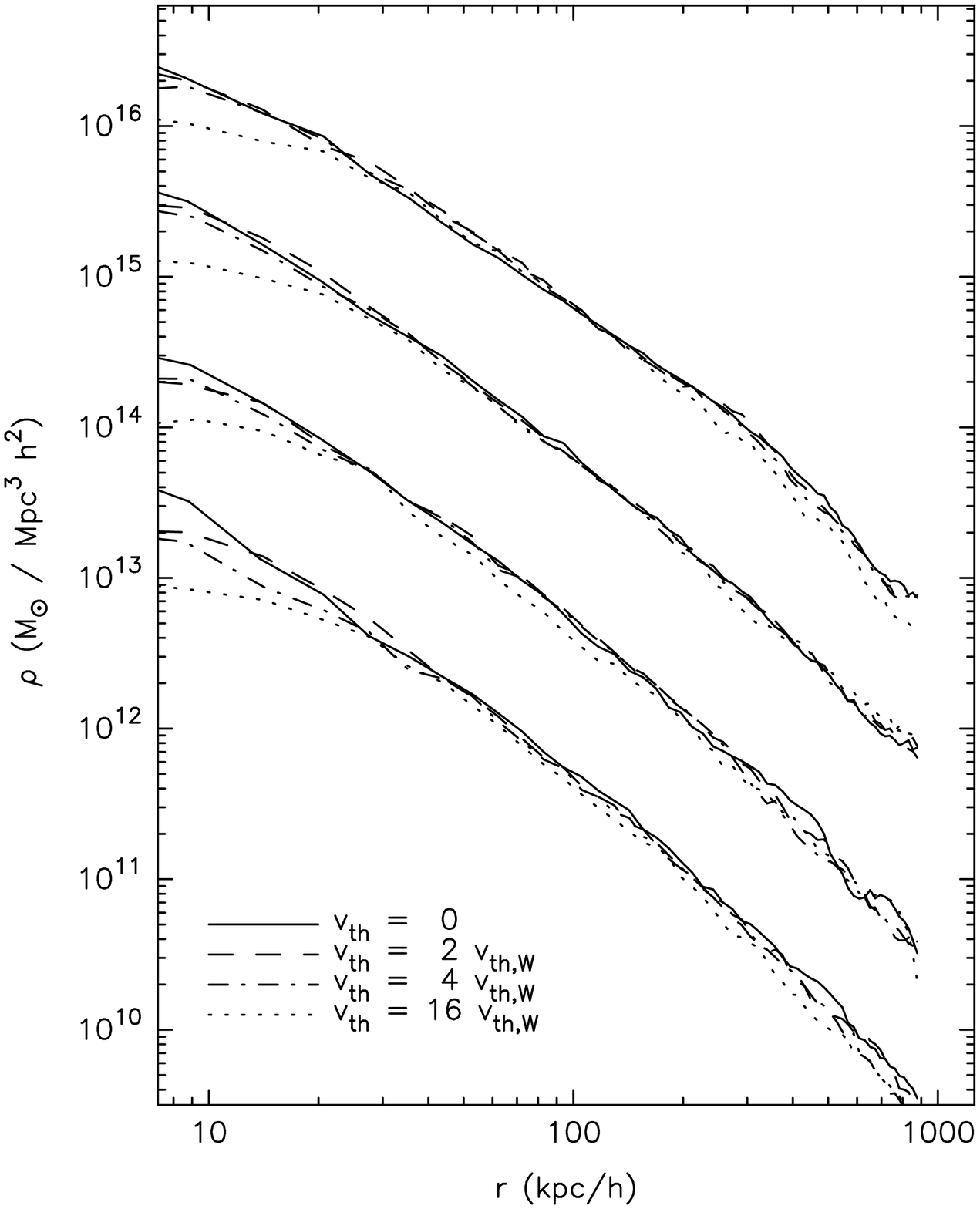}
\figcaption{Density profiles of the four host halos re-simulated
in the $\Lambda$WDM simulations with \fsl=1.7 \mpc\ and
\vth=0 (solid line, run \modVI, see also Fig. 2), \vth=2\vthw\ 
(dashed line, run \modVIII), \vth=4\vthw\ (dot-dashed line, run 
\modIX), and \vth=16\vthw\ (solid line, run \modX). The 
velocity \vthw\ is that corresponding to a warmon of 124 eV.
The profiles were shifted vertically by 1 in the log in order to 
avoid overlapping. Soft cores larger than $\sim0.01$ \rv\ appear 
only when the dispersion velocities are much larger than the 
warmon thermal velocities.}
\end{figure}

\begin{figure}
\plotone{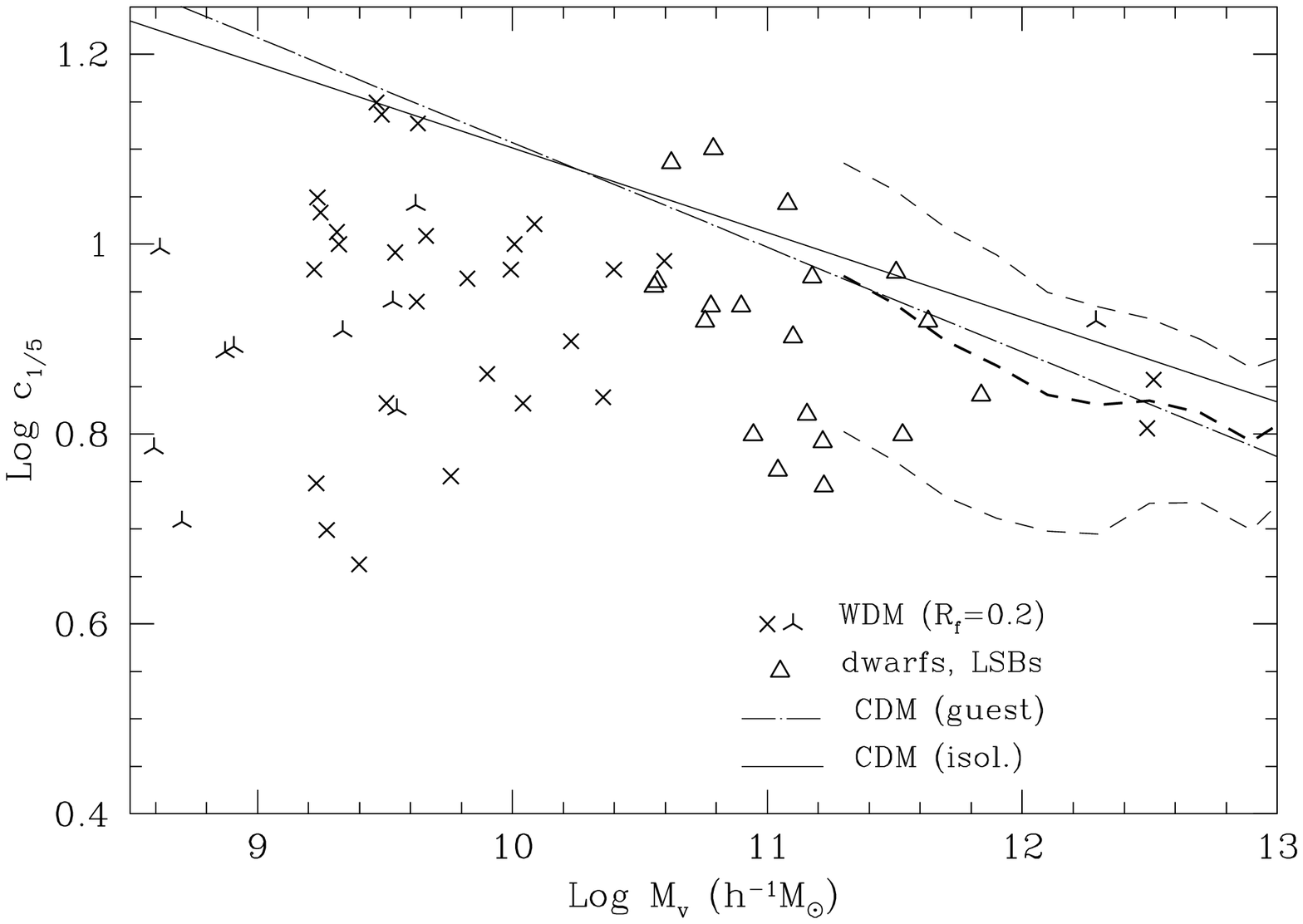}
\figcaption{Comparison of the c$_{1/5}$ concentration parameters
of \modII\ halos (\fsl=0.2 Mpc, crosses and skeletal triangles) with
those inferred from the rotation curves of dwarf and LSB galaxies (triangles).
Crosses correspond to a run presented in \Colinetal\ (L$_{\rm box}=15\mpch$, 
while skeletal triangles are from the run \modII\ presented here 
(L$_{\rm box}=7.5 \mpch$). The dashed lines are the average concentrations
and standard deviations of guest halos found in a $\Lambda$CDM simulation 
(Avila-Reese et al. 1999). The dot-dashed line is a linear fit to
these data, while solid line is the linear fit to data corresponding
to isolated halos in the same simulation. The observational data for
dwarf galaxies were taken from van den Bosch \& Swaters 2000 who
fit the halo component of their rotation curve decompositions to a
NFW profile finding this way c$_{\rm NFW}$. We pass from c$_{\rm NFW}$
to c$_{1/5}$ and take into account the difference in the definition
of virial radius of van den Bosch \& Swaters with our definition (
the difference in c$_{1/5}$ in not more than a factor of 1.05 larger
with our definition respecting that of these authors). The mass of the
halos is calculated from the V$_{200}$ reported by them:
$\mvir[\msunh]=4.2\times 10^5 V_{200}^3[\kms]$. We have used only those
galaxies for which a meaningful fit to the NFW is found. For the LSB
galaxies, data from van den Bosch et al. 2000 were used. The same procedure
described for the dwarf galaxies was used. We have considered only those
galaxies for which the estimated V$_{200}$ is smaller than the measured
V$_{\rm max}$. Unfortunately, WDM models and observations do not overlap 
too much. Nevertheless, one already sees that the halos of dwarf and LSB
galaxies are in average less concentrated than the CDM ones, being in 
better agreement with the WDM halos.}
\end{figure}

\begin{figure}
\plotone{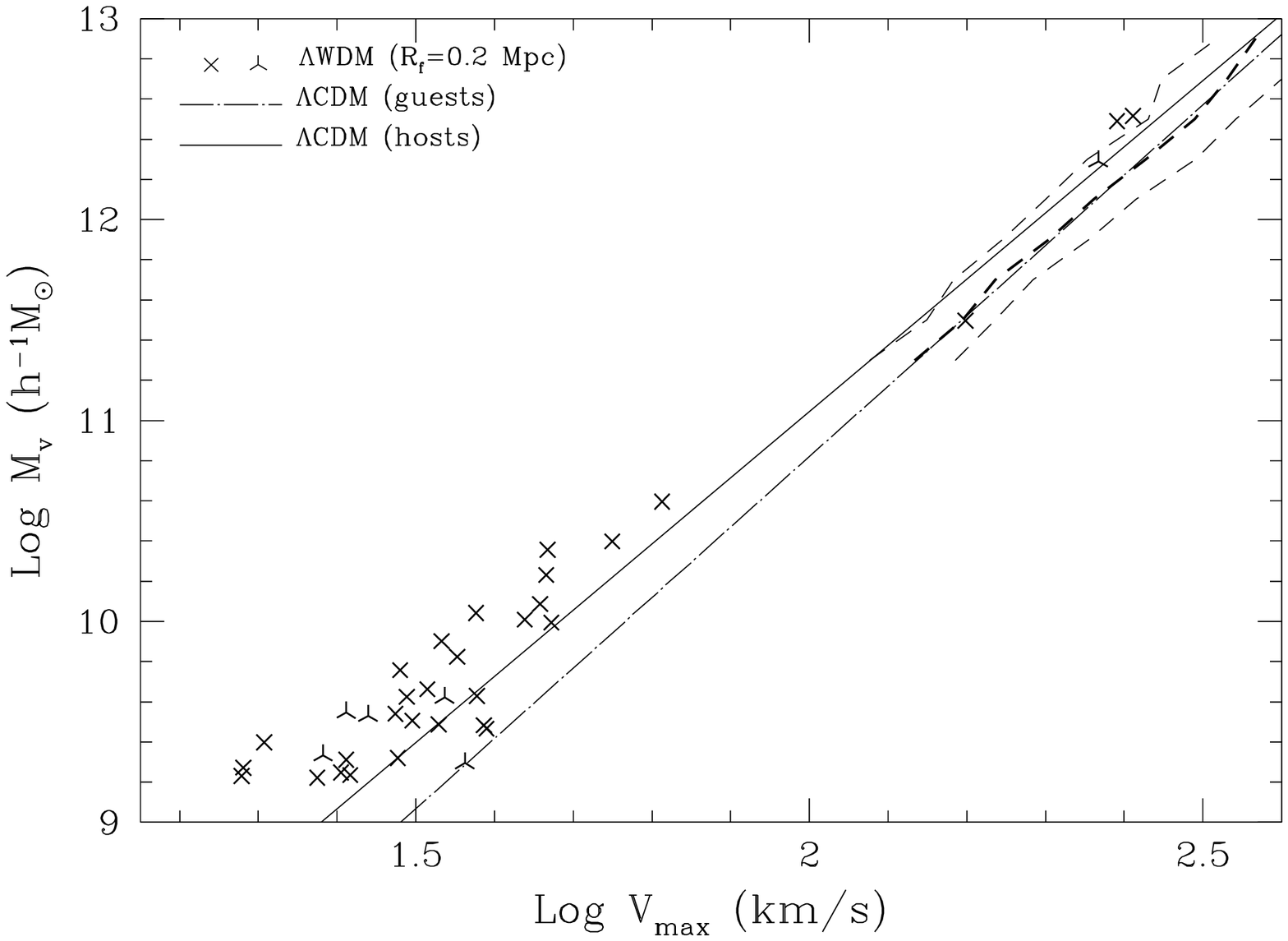}
\figcaption{Halo mass vs. its maximum circular velocity for 
the same runs presented in Fig. 4 and 6 ($\Lambda$WDM model with
\fsl=0.2 Mpc; crosses and skeletal triangles). For comparison, we
also plot the mass-velocity relation and its dispersion for
guest halos (dashed lines) found in a $\Lambda$CDM simulation, a linear fit 
to this relation (dot-dashed line), and a linear fit corresponding to 
isolated halos (solid line) (see Avila-Reese et al. 1999).}
\end{figure}

\end{document}